\documentclass[pdflatex,sn-mathphys-num]{sn-jnl}
\usepackage[T1]{fontenc}
\usepackage{graphicx}
\usepackage{amsmath,amssymb,amsfonts}
\usepackage{bm}
\usepackage{booktabs}

\graphicspath{{figures/}}

\newcommand{\M}{\mathcal M}
\newcommand{\ase}{\mathrm{as}}

\usepackage[textsize=footnotesize]{todonotes}
\usepackage[todos]{CMImacros}

\usepackage{xspace}
\newcommand{\hmol}{\ensuremath{\mathrm{H}_2}\xspace}

\hypersetup{
    pdftitle={Attosecond Wigner time delay of photoionization in the quantum critical systems and phase transitions},
    pdfauthor={Shuaidong Wang, Jiaji Zhang, Zhongchen Xing, Peixiang Lu, Min Li, Zheng Li}
}

\begin{document}

\title{Attosecond Wigner time delay of photoionization in the quantum critical systems and phase transitions}

\author[1]{\fnm{Shuaidong} \sur{Wang}}
\equalcont{These authors contributed equally to this work.}

\author*[2]{\fnm{Jiaji} \sur{Zhang}}
\email{zhang.jiaji.i92@kyoto-u.jp}
\equalcont{These authors contributed equally to this work.}

\author[3]{\fnm{Zhongchen} \sur{Xing}}
\author*[1]{\fnm{Peixiang} \sur{Lu}}
\email{lupeixiang@hust.edu.cn}
\author*[1]{\fnm{Min} \sur{Li}}
\email{mli@hust.edu.cn}
\author*[1,2]{\fnm{Zheng} \sur{Li}}
\email{zhengli@wit.edu.cn}

\affil[1]{School of Physics and Wuhan National Laboratory for Optoelectronics, Huazhong University of Science and Technology, Wuhan 430074, China}

\affil[2]{School of Optical Information and Energy Engineering, Wuhan Institute of Technology, Wuhan 430205, China}

\affil[3]{School of Physics, Peking University, Beijing 100871, China}

\abstract{%
The critical phenomena and phase transitions are profound concepts of many-body physics, yet their signatures in the photoelectron spectra remain largely unexplored. Here we report how the attosecond Wigner time delay in photoionization behaves in quantum critical systems, extending from stretched molecular hydrogen to periodic Hubbard model, as well as models of Bardeen-Cooper-Schrieffer (BCS) and high-temperature superconductors.
Based on theoretical analysis of these examples, we first prove that the Wigner time delay, i.e. the time the photoelectron takes to become a free electron, is a thermodynamic quantity. And the time delay exhibits distinctive characters in the crossover and phase transition of many-body systems, such as sign-flip and singularity at the critical points, implying drastic modulation of photoelectron velocity in the quantum systems undergoing phase transition. 
We showcase that the sign of Wigner time delay is relevant with the evolution of electron correlation in the phase transition and is topologically protected, as the flip of the photoionization delay sign is governed by a change in the winding number of the complex photoionization amplitude around the origin of the complex plane.
Our findings establish the attosecond photoionization time delay as a probe of quantum critical phenomena and the degree of electron correlation in the phase transitions, with applications to the study of strongly correlated systems such as the Mott insulator, metal-insulator transition (MIT) and superconductivity and ultrafast light-driven nonequilibrium phase dynamics.
}

\maketitle

Phase transition and attosecond photoionization time delay are traditionally understood as two unrelated theoretical concepts.
Understanding phases of matter and their transitions stands as a central pursuit in physical sciences, representing a continuously evolving frontier that drives fundamental discovery. 
Continuous phase transitions are often characterized by the evolution of an order parameter, accompanied by critical fluctuation, divergent correlation and universal scaling behaviour~\cite{Onsager1944,WilsonKogut1974}. Studies of phase transitions have revealed a wealth of collective phenomena, ranging from magnetic ordering and vortex unbinding in low-dimensional systems to superfluid–Mott-insulator transitions and topological phase transitions~\cite{KosterlitzThouless1973,Greiner2002,HasanKane2010,Cheng2026,Lin2026}. 
Beyond equilibrium, ultrafast optical excitation provides opportunities to manipulate electronic correlations and induce transient nonequilibrium states. For example, light-driven nonlinear lattice dynamics in YBa$_2$Cu$_3$O$_{6.5}$ has been shown to produce transient structural and electronic changes that favour superconductivity~\cite{mankowsky2014nonlinear}
In a seemingly distinct field, the Wigner time delay in photoionization has emerged as a sensitive probe of ultrafast electron dynamics~\cite{Wigner1955,Dahlstrom2012}. 
In atoms and molecules, photoionization time-delay studies have revealed orbital-dependent emission dynamics~\cite{Schultze2010,Klunder2011}, electron-correlation effects and ionization-pathway interference~\cite{Cirelli2018,Li2025,Mukherjee2024}, as well as resonance-induced electron dynamics and molecular electron localization~\cite{Huppert2016,Nandi2020,Vos2018}.
Photoemission time-delay measurements in condensed matter have further revealed electronic-state-dependent emission dynamics, electron transport, and screening effects~\cite{Cavalieri2007,Neppl2015,Locher2015,Potamianos2024,Heinrich2021}. 
Despite these advances, the connection between phase transitions and Wigner time delays, particularly how critical behavior manifests in the photoionization phases and time delays, remains largely unexplored and may further open a route towards using Wigner time delay to probe the ultrafast evolution of electronic correlations across light-driven nonequilibrium phase transitions.

In this work, We establish an exciting connection between attosecond photoionization time delay and the critical behavior of many-body systems.
We first examine this connection by analysing the Wigner time delay in the representative systems of crossover and phase transition, including stretched molecular hydrogen, periodic Hubbard model and superconductivity models, such as a 2D BCS model and the Hatsugai--Kohmoto (HK) model.
The stretched \hmol model first illustrate the sign of the Wigner delay, as a topological quantity, flips at the molecule-atom crossover, which is accompanied by the weak-strong electron correlation transition of Mott-Hubbard type. 
Analysis of the BCS and the HK wavefunctions reveals how electron correlation, thermal occupation and pairing affect the response of Wigner time delay in the thermodynamic limit, leaving signatures such as singularity and discontinuity at the critical point.
Our results ultimately establish a solid connection between the critical phenomena and Wigner time delay of photoionization, opening a route towards using attosecond metrology to probe critical behaviour in the many-body phase transitions.

\section*{Stretched \hmol}
At the outset, we start from molecular hydrogen with a stretched bond length, and calculate the photoionization Wigner time delays. 
It is helpful to treat the molecular hydrogen as a two-site half-filled Hubbard model in order to establish the correspondence with condensed matter systems, the evolution from weak to strong electron correlation is modulated by its bond length~\cite{Chiappe2007}. 
The Mott-Hubbard crossover of stretched \hmol from molecular to quasiatomic state captures the fundamental physics of phase transition in the strongly correlated systems~\cite{Chiappe2007,Kadzielawa2015}. 
In strongly correlated systems, such as the Mott insulators La$_2$CuO$_4$ and heavy-fermion compounds, the competition between electron localization and delocalization plays a central role in metal-insulator transition (MIT), and the ratio between the hopping energy $t$ and electron correlation $U$ dominates. The strong correlation arises from the Coulomb repulsion $U$ of electrons in the inner $d$ and $f$ orbitals, whose size is essentially smaller than the interatomic distance determined by the outer $s$ orbital of the metal atom. 
A minimal analogue of pressure-induced MIT is provided by the Mott-Hubbard crossover of \hmol, where bond stretching simulates the pressure effect and suppresses the interatomic hopping $t$ while the on-site repulsion $U$ remains nearly unchanged.
Electronic correlation arises from Coulomb interaction. In the weakly correlated regime, the electronic state is well approximated by a single Slater determinant, whereas strong correlation emerges when multiple electronic configurations become essential in the many-body wavefunction.
By elongating the bond length, the electron correlation in the \hmol evolves from the weakly to strongly correlated regime. Though a phase transition in the thermodynamic limit is not present, the molecule-atom crossover mimicks the metal-insulator transition, and we demonstrate that this Mott-Hubbard crossover reverses the sign of the Wigner delay in the vicinity of Cooper minimum, namely the photoelectron becomes faster than the free electron with the same momentum.

We now outline how the Wigner time delay is calculated from the complex photoionization amplitude $\M(E)$ for photoelectron with energy $E$\,. The amplitude includes electron removal and propagation in the continuum. The Wigner time delay is $\tau(E)=\hbar{\partial_E\arg \M(E) }\,.$
Physically, it characterizes the temporal shift of the outgoing photoelectron wavepacket relative to the free electron propagation, and its positive and negative values corresponds to delays and advances.
We choose the Wigner time delays near the Cooper minimum~\cite{Li2025}, where small perturbation in interfering ionization pathways to multiple electron configurations can produce rapid phase variations and the influence on the sign of $\tau(E)$ can be maximized. 

Since we are only interested in Hubbard physics for now, we map a soft-core potential Hamiltonian of stretched H$_2$ molecule to the corresponding two-site half-filled Hubbard model~\cite{Chiappe2007, AlvarezFernandez2002}, for which the Mott-Hubbard crossover occurs at the critical point $R_{\mathrm c}=4.5\,a_0$ and $U_{\mathrm c}=2\sqrt{3}t$ (see the details of the mapping in Section 1 of the Supplementary Information, SI).

\begin{figure}[tbp]
\centering
\includegraphics[width=.98\linewidth,keepaspectratio]{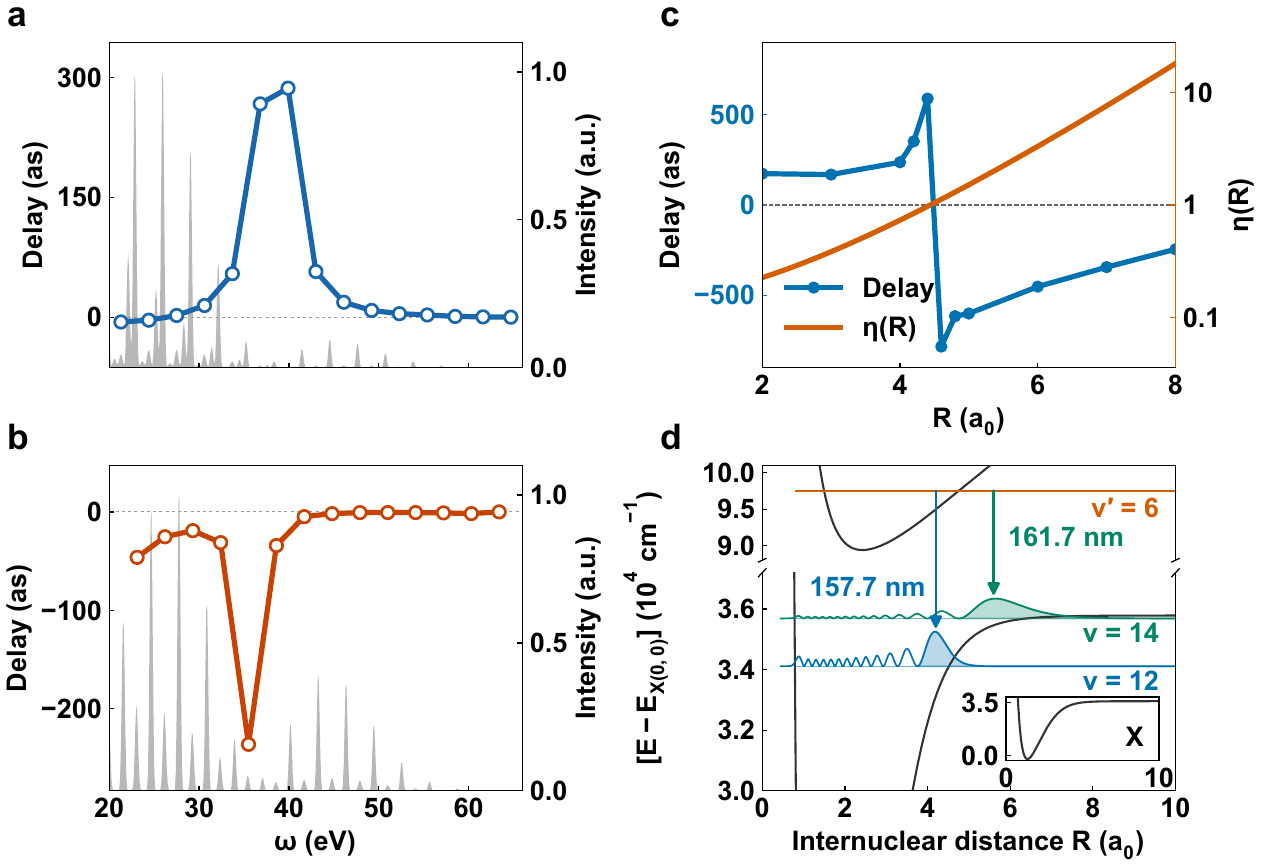}
\caption{\textbf{Molecular delay reversal upon bond stretching and a proposed vibrational preparation scheme.} \textbf{a\,, b}\,, RABBITT photoionization delays (coloured curve) and normalized photoelectron spectra (grey) from the two-electron TDSE for the ungerade ionic channel at fixed internuclear separations of \textbf{a}, $R=4\,a_0$\,, and \textbf{b}, $R=8\,a_0$\,. Pronounced positive and negative delay occur near Cooper-like minima in the photoelectron spectra. \textbf{c}, Evolution of the photoionization delay at the Cooper-like minimum (blue line) and the interaction-to-hopping ratio $\eta(R)=U/[2\sqrt{3}\,t(R)]$ (orange curve) with internuclear separation $R$\,. Here, $U$ and $t$ are the on-site Coulomb repulsion and inter-site hopping amplitude in the effective two-site half-filled Hubbard model. The condition $\eta=1$ defines the Mott–Hubbard crossover criterion, separating the hopping-dominated molecular regime ($\eta<1$) from the interaction-dominated atomic regime ($\eta>1$). \textbf{d}, Calculated potential-energy curves of the $X\,{}^1\Sigma_g^+$ and $B\,{}^1\Sigma_u^+$ electronic states, illustrating a proposed STIRAP scheme for transferring population from $X(\nu=0,J=0)$ to the highly excited vibrational states $X(\nu=12,14,J=0)$ to extended internuclear separations via the intermediate level $B(\nu'=6,J'=1)$\,. The blue and green curves represent the nuclear probability densities of the $\nu=12$ and $\nu=14$ states. Shaded regions highlight the largest integrated nuclear probability. The inset shows the full $X$-state potential well. All energies are referenced to $X(\nu=0,J=0)$.}
\label{fig:molecule}
\end{figure}

By solving the time-dependent Schr\"odinger equation (TDSE), we simulate attosecond metrology of the Wigner time delay with RABBITT (reconstruction of attosecond beating by interference of two-photon transitions) method~\cite{Paul2001, Agostini2024}.
For an \hmol molecule ionized by an attosecond extreme ultraviolet (XUV) pulse, the measured delay contains two contributions: the Wigner delay, $\tau_{\mathrm W}$, and the continuum--continuum (CC) delay, $\tau_{\mathrm{CC}}$,
$\tau_{\mathrm{RABBITT}}\simeq\tau_{\mathrm W}+\tau_{\mathrm{CC}}$~\cite{Klunder2011}. 
We focus on photoelectron kinetic energies between 20 and 60 eV, where the CC contribution is reduced compared with the low-energy regime~\cite{JiCC2024}. 
Figure~\ref{fig:molecule} reveals a pronounced flip of the Wigner time delay $\tau_{\mathrm E}$ associated with the ungerade ionic channel, in which the electron occupies the molecular orbit with odd parity: a positive feature of approximately $+230\,\ase$ at an internuclear separation of $R=4a_0$ changes to a negative feature of approximately $-240\,\ase$ at $R=8a_0$, where $a_0$ is the Bohr radius. 
The pronounced time delay variation occurs near a Cooper-like minimum in the H$_2$ photoelectron spectrum, where the destructive interference produces a rapid shift in the photoionization phase~\cite{Li2025, DellaPicca2009}. 
To understand the origin of the sign flip, we examine how bond stretching modifies the correlation of the electron wavefunction. 
We seek a simplified two-configuration electron wavefunction for the stretched H$_2$ of bond length $R$,
\begin{equation}
\ket{\Psi_{\mathrm H_2}(R)}=c_1(R)\ket{gg}+c_2(R)\ket{uu}\,,
\label{eq:h2ci}
\end{equation}
where $g$ and $u$ represent gerade and ungerade electron orbitals.
For $R<R_{\mathrm c}$, the wavefunction is dominated by the doubly occupied configuration $\ket{gg}$. 
When the bond is stretched, the contribution of the doubly occupied anti-bonding configuration $|uu\rangle$ progressively increases, reflecting the emergence of strong static electron correlation.
Crossing the Mott-Hubbard critical bond length, the two electrons become localized in individual atoms, which is equivalent to the growing multi-configuration character from the wavefunction viewpoint.

\begin{figure}[tbp]
\centering
\includegraphics[width=.90\linewidth,keepaspectratio]{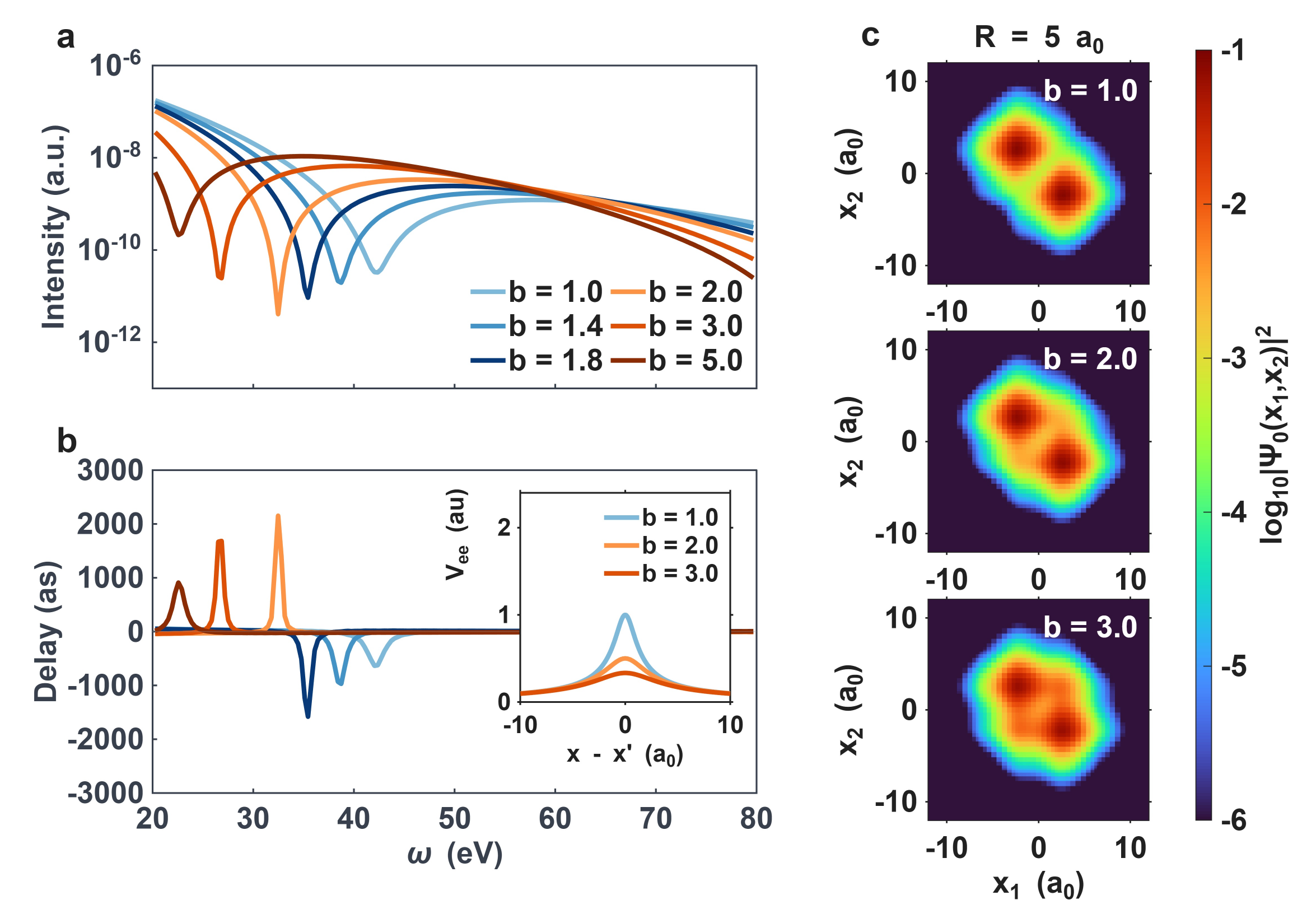}
\caption{\textbf{Interaction-controlled delay flip and initial state distribution \hmol with $R=5\,a_0$\,.} Photoelectron spectra (\textbf{a}) and time delays (\textbf{b}) as functions of photoelectron energy $\omega$ for different electron--electron (e-e) interaction strengths at a fixed internuclear separation $R=5\,a_0$\,. The interaction strength is controlled by the soft-core parameter $b$ through e-e repulsion potential
$
V_{\mathrm{ee}}(x_1,x_2;b)
=1/\sqrt{(x_1-x_2)^2+b^2}\,.
$ 
Increasing $b$ weakens the Coulomb repulsion and reduces electronic correlations, leading to a shift of the Cooper-like minimum and a reversal of delay from negative to positive between $b=1.8$ and $b=2.0$. The inset in \textbf{b} shows the e-e interaction potentials for different $b$.
\textbf{c\,,} Initial-state joint electron probability densities $|\Psi_0(x_1,x_2;b)|^2$ for $b=1.0$, $2.0$ and $3.0$. As $b$ increases, the probability density in the first and third quadrants ($x_1x_2>0$) increases, indicating an enhanced probability of both electrons occupying the same atomic site. This redistribution reflects a change from an interaction-dominated quasiatomic state towards a more hopping-dominated molecular state as electronic correlations weaken.}
\label{fig:bscan}
\end{figure}

We modulate the electron correlation in a more explicit way by artificially tuning the effective electron-electron (e-e) interaction strength in the parametrized soft-core potential model for a fixed bond length $R=5\,a_0$.
Figure~\ref{fig:bscan}(a) illustrate the shift of Cooper-like minimum when the e-e interaction is weakened by enhanced soft-core parameter $b$. 
Around the Cooper-like minimum, the time delays in Figure~\ref{fig:bscan}(b) flip the sign from negative to positive between the $b=1.8$ and $b=2.0$\,.

We briefly explain why the sign-flip point coincides with the critical point of phase transition in the stretched H$_{2}$ molecule.
Suppose that photoionization amplitude is in linear summation form, $\mathcal{M} = \sum_{m} \mathcal{M}_{m}$, and for \hmol molecule, $\M  = \mathcal{M}_{gg} + \mathcal{M}_{uu}\,$.
Each component $\mathcal{M}_{m}$ has distinct phase factor, namely $\mathcal{M}_{m} = A_{m} \, e^{i \theta_m}$.
Crossover or phase transition can be solely characterized by the one-particle Green's function, which can be expressed in Lehmann representation as an incoherent summation, $\mathcal{G} = \sum_{m} A_{m}^2$.
For Wigner time delay, we have ${\rm{arg}} \mathcal{M} = {\rm{tan^{-1}}} [\sum_{m} A_m \sin\theta_m/\sum_{m} A_{m} \cos\theta_m]$.
The critical points in Green's function $\mathcal{G}$ should arise from the sum of $A_{m}$.
These criticalities should be recovered in the Wigner time delay, as it also involves the sum of $A_{m}$ in the arguments.
The extra weighted factors $\theta_{m}$ do not remove the criticality.

To realize the proposed attosecond metrology probe of the Mott-Hubbard crossover of \hmol molecule, the stimulated Raman adiabatic passage (STIRAP) method can be used control the bond length of the diatomic molecule~\cite{Bigagli2024}, transferring the population from the initial $X\,{}^1\Sigma_g^+(v=0,J=0)$ state to the selected $X\,{}^1\Sigma_g^+(v=12,14,J=0)$ target states through an intermediate level $B\,{}^1\Sigma_u^+(v'=6,J'=1)$. 
STIRAP could prepare H$_2$ in the $X\,{}^1\Sigma_g^+(v=12,14,J=0)$ levels, providing access to extended internuclear separations for probing the predicted delay reversal 
For comparison with the fixed-$R$ results, we assign each target state a most probable distance $R_v^{\mathrm{peak}}=\mathrm{max}_R |\chi_v(R)|^2$, giving $4.180$, and $5.643$\,$a_0$ (see Fig. \ref{fig:molecule}d, and details in section 1 of the Supplementary Information). 

\section*{Periodic Hubbard Model}
\begin{figure}[tbp]
\centering
\includegraphics[width=.98\linewidth,keepaspectratio]{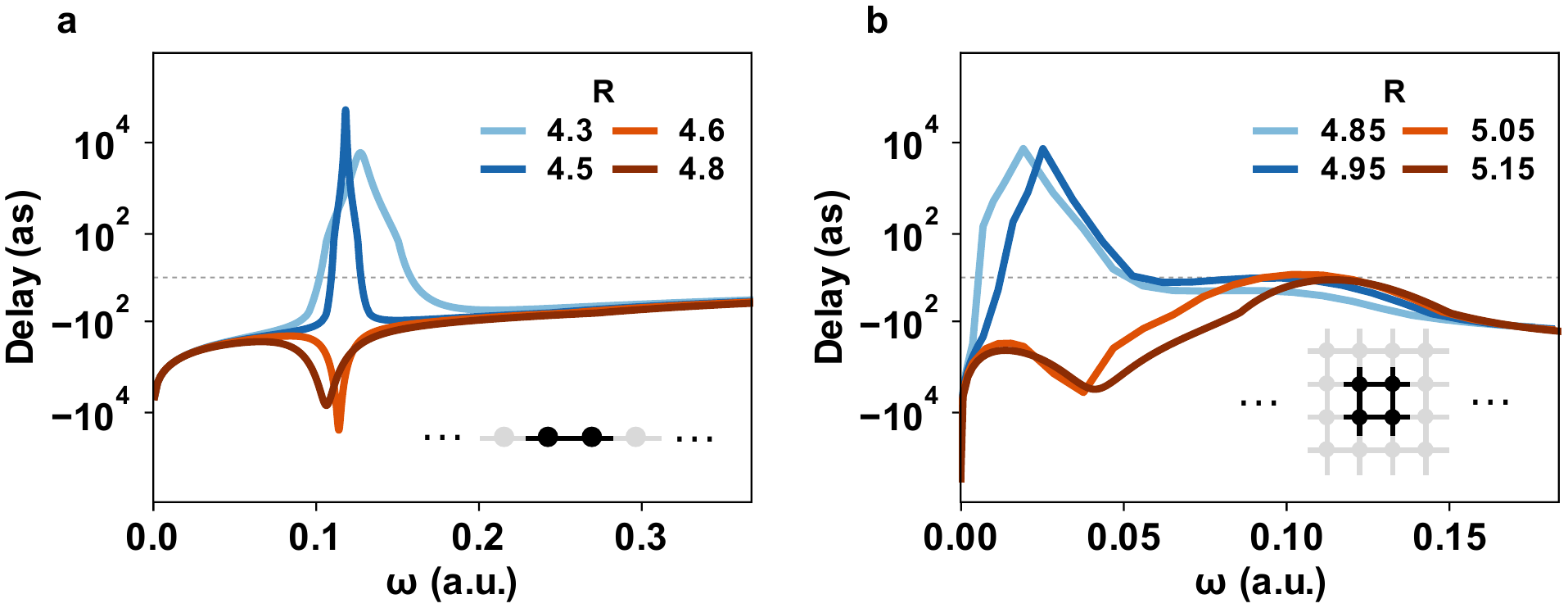}
\caption{\textbf{Delay reversal in finite periodic Hubbard lattices.} Channel-resolved Wigner time delays as functions of photoelectron energy ($\omega$) for \textbf{a}\,, a two-site Hubbard lattice and \textbf{b}\,, a four-site square Hubbard lattice with periodic boundary condition. In both systems, increasing inter-site separation $R$ induces the flip of the Wigner delay from positive to negative. The lower-right insets illustrate the corresponding Hubbard lattices, with black sites and bonds representing the finite lattice and grey sites and bonds indicating its periodic replicas. The complex emission amplitudes include Dyson orbital and photoelectron scattering from the residual $(N-1)$-electron system.}
\label{fig:lattice}
\end{figure}

Using exact diagonalization method, we extend our analysis from molecular \hmol to finite Hubbard lattices with periodic boundary condition.
We consider Hubbard Hamiltonian of a two-site and a four-site square unit with periodic boundaries~\cite{Hubbard1963, Danilov2022}. 
The $N$-electron initial state and the $m$-th $(N-1)$-electron state are used to construct the complex photoionization amplitudes for ionizing from the $j$-th Dyson orbital, $d_{mj}=\langle\Phi_m^{N-1}|c_j|\Psi_0^N\rangle$. 
Figure~\ref{fig:lattice} illustrates that the calculated Wigner time delays in the periodic lattices reverse sign as the inter-site distance $R$ crosses the critical value. 

\begin{figure}[tbp]
\centering
\includegraphics[width=.52\linewidth,keepaspectratio]{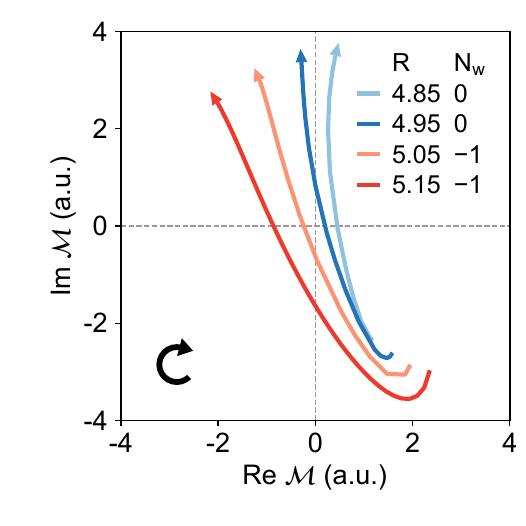}
\caption{\textbf{Complex photoionization amplitudes near a Cooper-like minimum.} Complex-plane trajectories of the photoionization amplitude $\M$ for the four-site square Hubbard model at different inter-site distances $R$\,. Arrows indicate the direction of increasing photoelectron energy, along which all trajectories rotate clockwise. The winding numbers $N_w$ are labeled for four trajectories, with $N_w=0$ corresponding to positive Wigner time delays and $N_w=-1$ to negative Wigner time delays.}
\label{fig:complex}
\end{figure}

The sign of the Wigner delay is topologically protected, because it is determined by the oriented winding number of the complex photoionization amplitude trajectories~\cite{Ji2024}. 
Trajectories of the periodic square Hubbard model in Fig.~\ref{fig:complex} provides a geometric interpretation of the delay reversal through their orientation and winding number around the origin of the complex plane. 
While the trajectories evolve clockwise (CW) with increasing photoelectron energy $E$, their winding numbers change with the inter-site separation $R$. For $R\leq4.98$, the winding number is zero and the delay is positive, whereas for $R\geq5.02$, the winding number is one and the delay is negative (see the relation of the winding number and Wigner delay in section 3 of the Supplementary Information). 

\section*{Scaling}
Now we proceed to the Wigner time delay of condensed matter systems in the thermodynamic limit.
Photoionization of a solid state system with $L$ sites or repetitive cells can be characterized by a complex amplitude $\M_L(\omega)$, where we work with \(\hbar=1\) and photoelectron energy $\omega$ for the convention of condensed matter physics.
We first demonstrate that the phase derivative of $\M_L$, i.e. the Wigner time delay, can maintain intensive under proper scaling condition, because excluding amplitude zeros and singular points, the size dependent $\M_L(\omega)$ factorizes as
\begin{equation}
\M_L(\omega)\propto L^\alpha
\left[A(\omega)+iB(\omega)\right],
\label{eq:delay_power}
\end{equation}
where $L$ denotes the system size, and $A(\omega)$ and $B(\omega)$ are real functions independent of $L$ (see detailed proof in the SI). 
Since the real-valued factor $L^\alpha$ changes only the magnitude of the transition amplitude, it does not contribute to the Wigner delay. 
More generally, we show that even when the phase of the photoionization amplitude scales with system size $L$, its size-dependent part is independent of energy, namely, $\M_L(\omega)\propto e^{i\gamma L+iF(\omega)}$, and the Wigner time delay retains the intensive nature, $\tau_L(\omega)=\partial_\omega F(\omega)=O(1)$.
An extensive contribution to the Wigner delay requires a phase that couples the photoelectron energy directly to the system size, e.g. in a form $\M_L(\omega)\propto e^{i\gamma\omega L+iF(\omega)}$, which leads to $\tau_L(\omega)=\gamma L+\partial_\omega F(\omega)$.
To exclude such cases, we showcase the structure of the photoionization amplitude, which involves a coherent superposition of electron-removal pathways, $\M_L(\omega)=\sum_n d_{n,L}(\omega)$,
with
\begin{equation}
d_{n,L}(\omega)=
\frac{
\langle n|\hat c_k|\psi_0\rangle
}{
\omega+E_n-E_0-\Omega_{\rm ph}+i\eta
},
\label{eq:delay_removal}
\end{equation}
where $|\psi_0\rangle$ and $|n\rangle$ are the initial many-body state and the photoionized states, the photon energy is $\Omega_{\rm ph}$. For an extensive energy-dependent phase to survive the coherent summation while the spectral density $G_L(\omega)=\sum_n|d_{n,L}(\omega)|^2$ maintains intensive, the individual contributions of $d_{n,L}(\omega)$ have to acquire a common phase proportional to $\omega L$. 
However, the system-size dependence enters through the many-body states and their transition matrix elements, whereas the photoelectron energy $\omega$ is an independent external parameter. The resulting scaling thus excludes the collective phase in the form $e^{i\gamma\omega L}$.
Provided \(A^2+B^2\neq0\) in Eq.~\ref{eq:delay_power}, the Wigner time delay $\tau_L(\omega)$ is intensive in the thermodynamic limit of infinitely many sites,
\begin{equation}
\begin{aligned}
\tau_L(\omega)
&
=\operatorname{Im}\frac{\partial_\omega \M_L(\omega)}{\M_L(\omega)}\,,
\\
\lim_{L\to{\infty}}\tau_L(\omega)
&=
\frac{A(\omega)\partial_\omega B(\omega)
-B(\omega)\partial_\omega A(\omega)}
{A^2(\omega)+B^2(\omega)}=O(1).
\end{aligned}
\label{eq:delay_intensive}
\end{equation}
We can now investigate the abundant characters of the Wigner time delay $\tau(\omega)$ for the phase transition of condensed matter systems, since $\tau(\omega)$ can serve as a new thermodynamic response function that manifest singularity and discontinuity near the critical points~\cite{budich2016}.

\section*{Superconductivity of Hatsugai--Kohmoto model}

The Hubbard model demonstrates the sign-flip of the correlation-dependent Wigner delay in the Mott--Hubbard crossover, but its rapidly growing Hilbert space prohibits analysis of Wigner delay in the thermodynamic limit. We turn to the Hatsugai--Kohmoto (HK) model with exactly solved wavefunctions~\cite{Hatsugai1992,Phillips2020,Mai2026}. The interaction in the HK model is local in momentum space, allowing the Hamiltonian to be decomposed into independent momentum blocks. 
The HK model is a prototypical system for the cuprate superconductors.
The Hamiltonian is characterized by a Coulomb repulsion $U$ and an attractive $s$-wave pairing interaction $g$, with an electronic bandwidth $W$ setting the energy scale~\cite{Phillips2020}. The HK pairing Hamiltonian in momentum space is written as
\begin{equation}
\hat{H}=\sum_k\left[\epsilon_k(\hat{n}_{k\uparrow}+\hat{n}_{k\downarrow})+U\hat{n}_{k\uparrow}\hat{n}_{k\downarrow}\right]-g\hat{\Delta}^{\dagger}\hat{\Delta},
\label{eq:hkham}
\end{equation}
with
$
\epsilon_k=-\frac{W}{2d}\sum_{\mu=1}^{d}\cos k_\mu, \hat{\Delta}=\sum_k\hat{b}_k, \hat{b}_k^\dagger=\hat{c}_{k\uparrow}^\dagger\hat{c}_{-k\downarrow}^\dagger.
$
Here, $\epsilon_k$ is the single-particle dispersion, $\hat{b}_k^\dagger$ is the $s$-wave pairing operator, and $d$ denotes the spatial dimension. Introducing the chemical potential and treating the pairing interaction at the mean-field level, the Hamiltonian contains only coupling between $k$ and $-k$ components,
$
\hat{H}_{\mathrm{MF}}=\sum_{k>0}\hat{H}_k,
$
with
\begin{equation}
\begin{aligned}
\hat{H}_k={}&\zeta_k\sum_{\sigma=\uparrow,\downarrow}(\hat{n}_{k\sigma}+\hat{n}_{-k\sigma})\\
&+U(\hat{n}_{k\uparrow}\hat{n}_{k\downarrow}+\hat{n}_{-k\uparrow}\hat{n}_{-k\downarrow})\\
&+\Delta^*(\hat{b}_k+\hat{b}_{-k})+\Delta(\hat{b}_k^\dagger+\hat{b}_{-k}^\dagger),
\end{aligned}
\label{eq:hkmeanfield}
\end{equation}
where $\zeta_k=\epsilon_k-\mu$ and $\Delta$ is the superconducting gap parameter. At half filling, the chemical potential is $\mu=U/2$. The summation runs over one half of the Brillouin zone, and each $(k,-k)$ block contains four fermionic modes, $(k\uparrow,k\downarrow,-k\uparrow,-k\downarrow)$, spanning a 16-dimensional Fock space. The eigenstates and eigenenergies can be obtained by Jordan--Wigner transformation and exact diagonalization.

In the weak pairing limit $g\ll U$, the superconducting gap parameter $\Delta$ is determined self-consistently by a BCS-type gap equation with effective density $N''(\omega)$~\cite{Phillips2020},
\begin{equation}
1=\frac{g}{2}\int d\omega\,\frac{N''(\omega)}{\sqrt{\omega^2+\Delta^2}}\,.
\label{eq:hkgap}
\end{equation}
The gap parameter can be approximated by
$\Delta\simeq(W-U)^{1/2}U^{1/2}e^{-W/(2g)}$,
for $0<U<W$. 
The superconductor pairing instability is governed by the gap parameter $\Delta$, which vanishes in the normal state with $\Delta=0$.

The exactly solved wavefunction of the HK model allows us to investigate how electron correlation, superconducting pairing, and temperature affect the photoionization Wigner delay without being restricted to finte size Hubbard clusters. 
The explicit wavefunction is essential because computing the Wigner delay requires access to the phase of the photoemission transition amplitude -- information that goes beyond what the single-particle Green's function alone can provide.
After obtaining the eigenstates and eigenenergies $|\psi_n^{(k)}\rangle$ and $E_n^{(k)}$ of each $(k,-k)$ block, we construct the photoionization amplitudes at finite temperatures as
\begin{equation}
\M_{T,k\sigma}(\omega)=\frac{1}{\sqrt{Z_k}}\sum_{n,m}\frac{e^{-\beta E_m^{(k)}/2}\langle\psi_n^{(k)}|\hat{c}_{k\sigma}|\psi_m^{(k)}\rangle}{\omega+E_n^{(k)}-E_m^{(k)}-\Omega_{\rm ph}+i\eta}.
\label{eq:hkfinite}
\end{equation}
Here, $m$ and $n$ denote the initial and final eigenstates of each momentum-pair block, respectively, and $Z_k=\sum_m e^{-\beta E_m^{(k)}}$ is the partition function, with $\beta=1/T$ and $k_B=1$, and $\eta>0$ is a small positive spectral broadening parameter. At finite temperature, the initial states enter the coherent transition amplitude with weights $e^{-\beta E_m^{(k)}/2}/\sqrt{Z_k}$.

The corresponding Wigner time delay is obtained from the energy derivative of the phase of the coherent photoionization amplitude,
\begin{equation}
\tau_{T,k\sigma}(\omega)=\frac{\partial}{\partial\omega}\arg\M_{T,k\sigma}(\omega)=\mathrm{Im}\left[\frac{\partial_\omega\M_{T,k\sigma}(\omega)}{\M_{T,k\sigma}(\omega)}\right].
\label{eq:hkdelay}
\end{equation}
Using Eq.~\eqref{eq:hkfinite}, the energy derivative of the transition amplitude is evaluated analytically as
\begin{equation}
\begin{aligned}
\partial_\omega\M_{T,k\sigma}(\omega)=-\frac{1}{\sqrt{Z_k}}\sum_{n,m}\frac{e^{-\beta E_m^{(k)}/2}\langle\psi_n^{(k)}|\hat{c}_{k\sigma}|\psi_m^{(k)}\rangle}{\left(\omega+E_n^{(k)}-E_m^{(k)}-\Omega_{\rm ph}+i\eta\right)^2}.
\end{aligned}
\label{eq:hkderivative}
\end{equation}
At finite temperature, the superconducting gap $\Delta(T)$ modifies the eigenstates in each $(k,-k)$ block, and the corresponding Wigner delay is determined by the coherent summation of the thermally weighted photoionization amplitudes and the Wigner delay.

\begin{figure}[tbp]
\centering
\includegraphics[width=.90\linewidth,keepaspectratio]{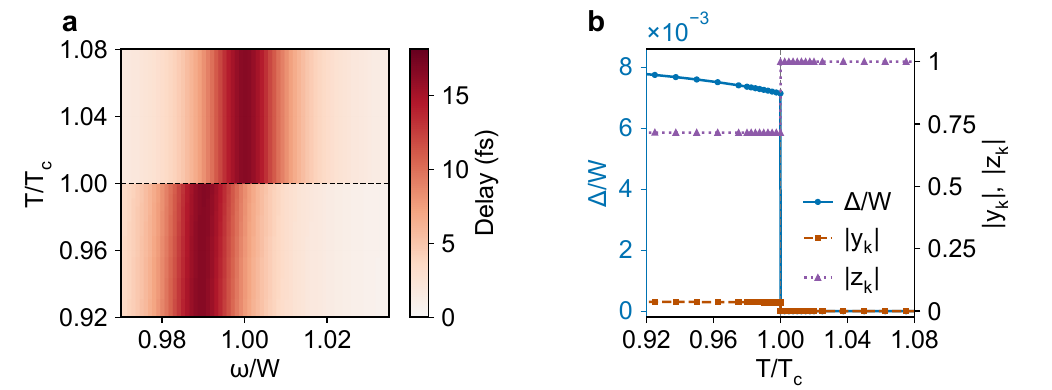}
\caption{\textbf{Temperature-dependent Wigner delays in the HK model.} \textbf{a}, Finite-temperature Wigner delay at $U/W=0.1$ and $\epsilon_k=\mu$, with $\Delta(T)$ determined by self-consistent thermodynamics. The horizontal dashed lines indicate the equilibrium superconducting transition $T=T_c$. \textbf{b}, Temperature dependence of the self-consistent pairing amplitude \(\Delta/W\) (blue, left axis) and  the magnitudes of the two-pair and one-pair spin-singlet coefficients, $|y_k|$ and $|z_k|$ (right axis)\,.
the coefficients are obtained from the superconducting ground state of the $(k,-k)$ block, as defined in Eq.~\ref{eq:hk_wavefunction}\,. All quantities exhibit discontinuous changes at $T=T_c$\,.}
\label{fig:hk}
\end{figure}

Figure~\ref{fig:hk}(a) the temperature dependence of the Wigner delay across the superconductivity phase transition for the electron correlation degree $U/W=0.1$ and pairing strength $g/W=0.1$.
A pronounced Wigner delay peak appears near $\omega/W=1$, with a discontinuity in the vicinity of the critical temperature $T_c$.
To understand this behavior, we consider the variational ground-state wavefunction of the HK model \cite{Phillips2020},
\begin{equation}
|\psi\rangle =
\prod_{k>0}
\left[
x_k+y_k b_k^\dagger b_{-k}^\dagger
+\frac{z_k}{\sqrt{2}}
\left(b_k^\dagger+b_{-k}^\dagger\right)
\right]|0\rangle,
\label{eq:hk_wavefunction}
\end{equation}
where $b_k^\dagger=c_{k\uparrow}^\dagger c_{-k\downarrow}^\dagger$ creates a Cooper pair, and $x_k$, $y_k$ and $z_k$ characterize the empty, two-pair and one-pair spin-singlet configurations, respectively, with $|x_k|^2+|y_k|^2+|z_k|^2=1$. Figure~\ref{fig:hk}(b) shows that the superconducting gap $\Delta$ and the configuration amplitudes $|y_k|$ and $|z_k|$ exhibit discontinuous changes at $T_c$, revealing an abrupt redistribution of the electronic configurations.
As the temperature approaches $T_c$, the superconducting gap closes, and the delay peak moves toward the normal-state position $\omega/W=1$. 
The discontinuity of $\tau_T(\omega)$ demonstrates the nature of Wigner delay as a new thermodynamic response function in the probe of phase transition.

\section*{Topological phase transition of BCS superconductor}
We push our investigation into the Wigner delays in varieties of phase transition, such as the topological phase transition.
The two-dimensional BCS model with $p_x+ip_y$ pairing has a solvable multi-reference wavefunction. Unlike the independent momentum blocks of the HK model, particle-number projection connects pairing configurations of different momenta while fixing the number of Cooper pairs. A macroscopic number of configurations contribute coherently to the transition amplitude. The model Hamiltonian is~\cite{iba2009,rombouts2010}
\begin{equation}
\begin{aligned}
\hat H={}&\sum_{\bm k}\frac{\bm k^2}{2m}\hat n_{\bm k}
-\frac{G}{4m}\sum_{\bm k\ne\bm k'}(k_x-ik_y)(k_x'+ik_y')
\hat c_{\bm k}^{\dagger}\hat c_{-\bm k}^{\dagger}
\hat c_{-\bm k'}\hat c_{\bm k'},
\end{aligned}
\label{eq:bcs_hamiltonian}
\end{equation}
where $G>0$ is the coupling constant and $m$ is the fermion mass. Within the mean-field treatment, the Bogoliubov transformation gives the quasiparticle energy
\begin{equation}
E_{\bm k}=\frac12\sqrt{(\bm k^2-\mu)^2+\bm k^2\Delta^2},
\label{eq:bcs_energy}
\end{equation}
for $m=1$. Here $\mu$ is the chemical potential and $\Delta$ is the gap parameter, determined by the self-consistent gap equations (see Section 6 of the Supplementary Information). In the reduced Brillouin zone $\bm\Omega^+$ of $k_x>0$, we choose $L$ momentum pairs, $M$ Cooper pairs, the filling fraction $x=M/L$ and the rescaled coupling $g=GL$. The Read--Green (RG) separating the phases are given by $\mu=0$ ($2x=1-1/g$)~\cite{iba2009}.
The RG line marks the topological quantum phase transition between the strong--pairing and weak--pairing phase where the bulk gap closes and the winding number $w$ jumps from $w=0$ (strong--pairing) to $w=1$ (weak--pairing), which is a topological superconductor.

The multi-reference ground state wavefunction is defined by projecting the Bogoliubov vacuum $|0_\gamma\rangle$ to a fixed number of $M$ Cooper pairs,
\begin{equation}
|\Psi_M\rangle=\mathcal N_M\int_0^{2\pi}\frac{d\theta}{2\pi}
 e^{i\theta(\hat N-2M)}|0_\gamma\rangle,
\label{eq:bcs_ground}
\end{equation}
where $\hat N$ is the particle-number operator and $\mathcal N_M$ is the normalization factor. The projected state contains a coherent sum of pairing configurations with the same particle number. Ionizing an electron with momentum $\bm k$ connects this initial state to the Bogoliubov quasiparticle occupation states $|\bm n\rangle$. The complex photoionization amplitude is
\begin{equation}
\M_{\bm k}(\omega)=\sum_{\bm n}
\frac{\langle\bm n|\hat c_{\bm k}|\Psi_M\rangle}
{\omega+E_{\bm n}-E_M-\Omega_{ph}+i\eta},
\label{eq:bcs_amplitude}
\end{equation}
where $E_{\bm n}$ and $E_M$ are the energies of the final configurations and the initial state.
Ionizing a photoelectron with momentum $\bm k$ leaves one Bogoliubov quasiparticle at $-\bm k$. Each remaining pair with momentum $(\bm p,-\bm p)$ can have either zero or two quasiparticle excitations, with coefficients
\begin{equation}
\begin{aligned}
A_{\bm p}(z)&=|u_{\bm p}|^2+|v_{\bm p}|^2z,\\
B_{\bm p}(z)&=u_{\bm p}v_{\bm p}(z-1),\qquad z=e^{2i\theta}.
\end{aligned}
\label{eq:bcs_pair_coefficients}
\end{equation}
The Fourier representation of the energy denominator introduces the time $t$, and each excited momentum pair contributes a phase $e^{2iE_{\bm p}t}$\,. The transition amplitude becomes
\begin{equation}
\begin{aligned}
\M_{\bm k}(\omega)={}&-i\mathcal N_M v_{\bm k}
\int_0^\infty dt\,
 e^{i(\omega+E_0+E_{\bm k}-E_M+i\eta)t}\\
&\times\oint\frac{dz}{4\pi i}\,z^{-M}
\prod_{\substack{\bm p\in\bm\Omega^+\\\bm p\ne\bm k}}
\left[A_{\bm p}(z)+B_{\bm p}(z)e^{2iE_{\bm p}t}\right],
\end{aligned}
\label{eq:bcs_integral}
\end{equation}
Here $E_0$ is the Bogoliubov vacuum energy, and the contour integral accounts for particle-number projection. The product sums the contributions from all other momentum pairs, including the phases $e^{2iE_{\bm p}t}$ associated with quasiparticle excitations.

In the thermodynamic limit, $L,M\to\infty$ at filling fraction $x$, the momentum sum becomes an integral over the single-particle energy $\varepsilon=k^2/2$, with the two-dimensional density of states $\rho(\varepsilon)=1/\Omega_c$ for $0<\varepsilon<\Omega_c$. The other momentum pairs product in Eq.~\eqref{eq:bcs_integral} takes the form $e^{L\Phi(z,t)}$, where
\begin{equation}
\begin{aligned}
\Phi(z,t)={}&\int_0^{\Omega_c}d\varepsilon\,\rho(\varepsilon)
\int_{-\pi/2}^{\pi/2}\frac{d\phi}{\pi}\\[-2pt]
&\times\ln\!\left[A(\varepsilon,z)
+e^{2iE(\varepsilon)t}e^{i\phi}B(\varepsilon,z)\right].
\end{aligned}
\label{eq:bcs_phi}
\end{equation}
In the thermodynamic limit, the product over momentum pairs takes the form $e^{L\Phi(z,t)}$. Together with the particle-number projection factor $z^{-M}$ and the time-dependent phase, the transition amplitude is dominated by the effective action
\begin{equation}
\mathcal S(z,t)
=i(\omega+E_0+E_{\bm k}-E_M+i\eta)t
+L[\Phi(z,t)-x\ln z]
\,.
\label{eq:bcs_action}
\end{equation}
The saddle point $z_0(t)$ and the saddle point of time $t_0$ are determined by
\begin{equation}
\begin{aligned}
\Phi_z(z_0,t_0)&=\frac{x}{z_0},\\
\Phi_t(z_0,t_0)&=-i\frac{\omega+E_0+E_{\bm k}-E_M+i\eta}{L}.
\end{aligned}
\label{eq:bcs_saddle}
\end{equation}
For a regular saddle point, the Wigner delay in the thermodynamic limit is (see proof in Section 6 of the Supplementary Information)
\begin{equation}
\tau_{\bm k}^{(\infty)}(\omega)=\operatorname{Re}t_0,
\label{eq:bcs_limit}
\end{equation}
where $\tau_{\bm k}^{(\infty)}(\omega)$ is the Wigner delay for $L\to\infty$, and $\partial_\omega\operatorname{Im}\mathcal S(t_0)=\operatorname{Re}t_0$. 
The Wigner delay near the Read--Green lines depends on the changes in the saddle time $t_0$ that follows the filling $x$ and the running coupling strength $g$. 
Two characters can be identified -- a cusp associated with merging saddle points and a discontinuous jump associated with competing saddle point contributions.

For photoelectron energy and momentum $\omega$ and $\bm k$, the pairing parameters vary along a path $\xi$ in phase diagram, with $x=x(\xi)$ and $g=g(\xi)$. 
The saddle time satisfies
\begin{equation}
f'(t_0;\xi)+i\mathcal E_\infty(\xi)=0,
\qquad
\mathcal E_\infty=-\mu x-\frac{\Delta^2}{4g},
\label{eq:bcs_energy_limit}
\end{equation}
where $f(t;\xi)=\Phi(z_0(t;\xi),t;\xi)-x(\xi)\ln z_0(t;\xi)$.
\begin{figure}[tbp]
\centering
\includegraphics[width=.90\linewidth,keepaspectratio]{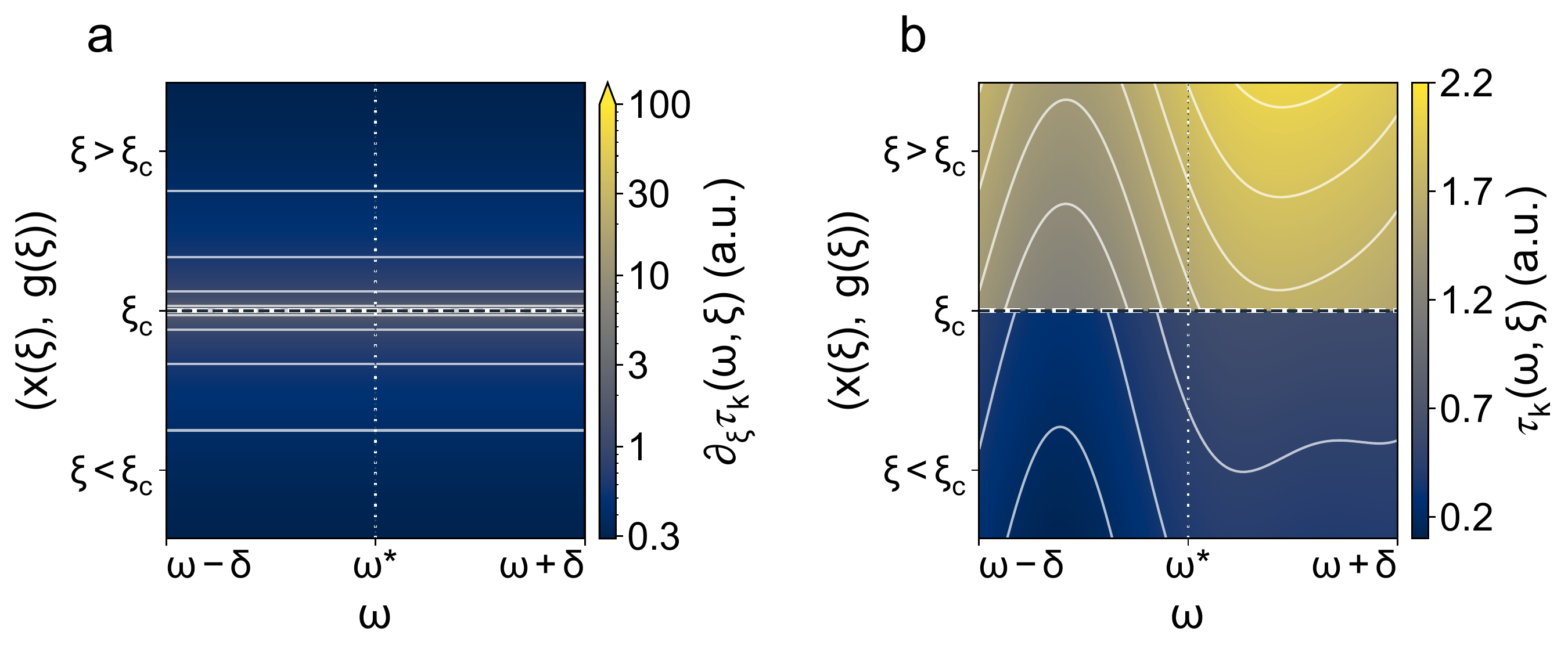}
\caption{\textbf{Schematic illustration of two behaviors of the Wigner time delay near the Read–Green (RG) phase boundaries.} The filling fraction $x$ and rescaled coupling $g$ vary along a path $\xi$ across the RG line at $\xi_c$. \textbf{a}\,, Cusp associated with the vanishing curvature of the effective action.
The Wigner delay remains finite, while its derivative becomes singular, $\partial_\xi\tau_{\boldsymbol{k}}(\omega,\xi)\propto|\xi-\xi_c|^{-1/2}$. 
\textbf{b}\,, Discontinuous jump of the Wigner delay $\tau_{\boldsymbol{k}}(\omega,\xi)$, arising from the competition between two saddle points. The exchange of their dominant contributions to the coherent photoionization amplitude leads to a finite jump in the thermodynamic limit. $\omega^*$ denotes the critical photoelectron energy by the corresponding saddle-point conditions, at which the cusp singularity or discontinuous jump can occur. The quantities $\omega^*-\delta$ and $\omega^*+\delta$ define a narrow photoelectron energy window around $\omega^*$\,. The horizontal dashed lines mark the critical point $xi_c$\,.}
\label{fig:BCS}
\end{figure}
A cusp occurs when the curvature vanishes, $f''(t_c;\xi_c)=0$. 
With $f'''(t_c;\xi_c)\neq 0$ and a non-zero response along the path, 
he derivative of Wigner delay exhibit a critical exponent of $-\frac12$ that leads to a singular slope (see Figure \ref{fig:BCS}a), 
\begin{equation}
\frac{\partial\tau_{\bm k}^{(\infty)}}{\partial x} 
= \frac{\partial\tau_{\bm k}^{(\infty)}}{\partial \xi} 
\frac{\partial \xi}{\partial x} 
\propto|\xi-\xi_c|^{-1/2}.
\end{equation}
The critical phenomena appear for $g(\xi)$.

A discontinuous jump arises from the competition between two regular saddle points, as their contributions enter the transition amplitude coherently, $\M_{\bm k}\simeq\M_{\bm k,1}+\M_{\bm k,2}$, giving (see proof in Section 6 of the Supplementary Information)
\begin{equation}
\tau_{\bm k}\simeq\operatorname{Re}\!\left[
\frac{t_{0,a}\M_{\bm k,a}+t_{0,b}\M_{\bm k,b}}
{\M_{\bm k,a}+\M_{\bm k,b}}\right].
\label{eq:bcs_two_saddles}
\end{equation}

Here $t_{0,a}$ and $t_{0,b}$ are the complex stationary times associated with the two saddle points. When either contribution dominates, the Wigner delay approaches the real part of the corresponding saddle time. To illustrate how their competition produces a discontinuity, we consider the relative amplitude

\begin{equation}
\frac{\M_{\bm k,b}}{\M_{\bm k,a}}
=e^{L\alpha(\xi-\xi_c)},\qquad \alpha>0,
\label{eq:bcs_saddle_ratio}
\end{equation}

where $\xi$ is a control parameter and $\xi_c$ denotes the point at which the two saddle contributions exchange dominance. Assuming that the two contributions have the same phase at the selected photoelectron energy, Eq.~\eqref{eq:bcs_two_saddles} becomes

\begin{equation}
\tau_{\bm k}(\xi)\simeq
\frac{
\operatorname{Re}t_{0,a}
+e^{L\alpha(\xi-\xi_c)}\operatorname{Re}t_{0,b}
}{
1+e^{L\alpha(\xi-\xi_c)}
}.
\label{eq:bcs_jump_crossover}
\end{equation}

For finite $L$, the delay evolves smoothly between the two saddle times. In the thermodynamic limit, the relative amplitude approaches

\begin{equation}
\lim_{L\to\infty}e^{L\alpha(\xi-\xi_c)}
=
\begin{cases}
0, & \xi<\xi_c,\\
+\infty, & \xi>\xi_c.
\end{cases}
\label{eq:bcs_saddle_limit}
\end{equation}

Consequently, the dominant saddle switches abruptly at $\xi_c$, and the Wigner delay approaches $\operatorname{Re}t_{0,a}$ and $\operatorname{Re}t_{0,b}$ on opposite sides of the transition. When the real parts of the saddle actions cross and the corresponding saddle times differ, this switch produces a discontinuous jump,

\begin{equation}
\Delta\tau_{\bm k}^{(\infty)}
=\operatorname{Re}
\left[t_{0,b}(\xi_c)-t_{0,a}(\xi_c)\right].
\label{eq:bcs_jump}
\end{equation}

The cusp and jump are two distinct responses of the coherent configuration sum. Their occurrence near the Read--Green line is determined by the saddle-point structure.

\section*{Conclusion}

By extending our investigation from molecular systems to periodic Hubbard lattices and as well as the HK and BCS model, we have established a connection between the critical behaviour of many-body systems and the Wigner delay of the photoionization. 
Abrupt reorganization of electronic configurations contribute coherently in photoionization pathways near the critical point, reshaping the interference and producing pronounced variations in the Wigner time delay.
In \hmol and finite periodic Hubbard lattices, the Wigner delay reverses sign as the internuclear or inter-site distance goes through the critical bond length of the Mott-Hubbard crossover. In the HK model, superconducting pairing and thermal excitation modify the Wigner delay through coherent interference of the ionization transition amplitudes. 
In the topological phase transition of BCS model, the Wigner delay is determined by the real part of the stationary time for a regular saddle point. The vanishing curvature and the competition between different saddle contributions can lead to a cusp and a discontinuous jump in the Wigner delay.
The Wigner delay is found to be a new response function of condensed matter systems in the thermodynamic limit. 
Our findings establish the connection between attosecond photoionization delays and the critical phenomena of many-body system, providing a route to the ultrafast probe of phase transition of transient non-equilibrium phases using attosecond metrology~\cite{mankowsky2014nonlinear,Cavalieri2007}.

\section*{Methods}
\subsection*{One-dimensional two-electron TDSE}
We solve the two-electron time-dependent Schr\"odinger equation for a one-dimensional diatomic model with fixed nuclei separated by $R$. Both electrons move along the molecular axis and interact through soft-core Coulomb potentials. The wavefunction is represented using a finite-element discrete-variable representation and propagated using a Lanczos method. For each internuclear distance, the correlated ground state is obtained by imaginary-time propagation and subsequently driven by an attosecond XUV pulse train and an infrared (IR) dressing field, both linearly polarized along the molecular axis. The laser interaction is treated in the velocity gauge. We implement the RABBITT scheme using an 800-nm IR field and an XUV pulse train centred on its 35th harmonic. By scanning the relative XUV--IR delay, we resolve the photoelectron sideband oscillations and extract their $2\omega_{\mathrm{IR}}$ phase to determine the channel-resolved RABBITT delay. The model Hamiltonian, laser pulse parameters and numerical implementation are detailed in Section 1 of the Supplementary Information.
For the proposed STIRAP preparation, we calculate the $X\,{}^1\Sigma_g^+$ and $B\,{}^1\Sigma_u^+$ potential-energy curves using full configuration interaction with the aug-cc-pVTZ basis in PySCF, solve the nuclear Schrödinger equation including rotation using a sine discrete-variable representation, and obtain the pump and Stokes wavelengths from the rovibrational energy differences connecting $X(v=0,J=0)$ and $X(v=12,14,J=0)$ to the common intermediate level $B(v'=6,J'=1)$.

\subsection*{Wigner delay of periodic Hubbard lattice}
We calculate photoionization delays for $2\times1$ and $2\times2$ Hubbard lattices with periodic boundary conditions~\cite{Hubbard1963}. Exact diagonalization in occupation-number bases yields the initial $N$-electron state and the residual $(N-1)$-electron eigenstates. We obtain the complex photoionization amplitude $\M_c(E)$ by solving the coupled-channel scattering problem for the photoelectron interacting with the residual system. The same detected final-state channel amplitudes are summed coherently. The Wigner delay is evaluated as $\tau_c(E)=\hbar\,\partial_E\arg\M_c(E)$, where $E$ is the photoelectron kinetic energy.

\section*{Acknowledgement}
We thank Phillip Werner, Lei Geng, Jiabao Ji for helpful discussions.
This study is supported by the National Key R\&D Program (Grant No.~2023YFA1406801, and the National Natural Science Foundation of China (Grant No. 12674326, 12234002, 92250303, 12434010).

\section*{Data availability}
The data supporting the findings of this study are available from the corresponding author upon reasonable request.

\bibliography{main}

\end{document}


\title{Attosecond Wigner time delay of photoionization in the quantum critical systems and phase transitions}
\subtitle{Supplementary Information}

\author[1]{\fnm{Shuaidong} \sur{Wang}}
\equalcont{These authors contributed equally to this work.}

\author*[2]{\fnm{Jiaji} \sur{Zhang}}
\email{zhang.jiaji.i92@kyoto-u.jp}
\equalcont{These authors contributed equally to this work.}

\author[3]{\fnm{Zhongchen} \sur{Xing}}
\author*[1]{\fnm{Peixiang} \sur{Lu}}
\email{lupeixiang@hust.edu.cn}
\author*[1]{\fnm{Min} \sur{Li}}
\email{mli@hust.edu.cn}
\author*[1,2]{\fnm{Zheng} \sur{Li}}
\email{zhengli@wit.edu.cn}

\affil[1]{School of Physics and Wuhan National Laboratory for Optoelectronics, Huazhong University of Science and Technology, Wuhan 430074, China}

\affil[2]{School of Optical Information and Energy Engineering, Wuhan Institute of Technology, Wuhan 430205, China}

\affil[3]{School of Physics, Peking University, Beijing 100871, China}


\maketitle

\noindent This Supplementary Information presents the two-electron H$_2$ TDSE calculations, the periodic small-lattice Hubbard models, the real-valued size-prefactor analysis of the Wigner delay, superconducting HK momentum blocks at zero and finite temperature, and the full number-projected chiral $p$-wave BCS derivation. The HK and BCS sections use coherent transition amplitudes, number projection and thermodynamic saddle-point methods. The discussion of electron-correlation-sensitive attosecond delays is motivated by recent interferometric experiments \cite{Li2025} and molecular ionization-pathway theory \cite{Mukherjee2024}.

\section{Two-electron \hmol TDSE and interaction control}
\subsection{Hamiltonian, parameters and nuclear configuration}
For electron coordinates $x$ and $x'$ along the molecular axis and fixed protons at $\pm R/2$, the field-free Hamiltonian is
\begin{align}
H_0={}&-\frac12\big(\partial_x^2+\partial_{x'}^2\big)
-\sum_{s=\pm1}\left[\frac{1}{\sqrt{(x-sR/2)^2+a^2}}+
\frac{1}{\sqrt{(x'-sR/2)^2+a^2}}\right] \notag\\
&+\frac{1}{\sqrt{(x-x')^2+b^2}}.
\label{eq:tdse}
\end{align}
All quantities are expressed in atomic units unless stated otherwise. Here $a$ and $b$ are the electron--nucleus and electron--electron soft-core parameters, respectively. The reference potential uses $a=b=1$; in the fixed-geometry scan, $a=1$ and $b$ is varied. The interelectronic separation is $x-x'$: the electron--electron term must therefore depend on $(x-x')^2+b^2$, not on $x^2+b^2$ with a single electron coordinate. The electron--electron potential becomes weaker as $b$ increases, while the nuclear positions and electron--nucleus potential are unchanged. The laser interaction in length gauge takes the form $H_{\mathrm{int}}(t)=(x+x')F(t)$ under the corresponding field-sign convention.

Two sets of TDSE calculations are discussed. The internuclear-distance scan compares $R=4a_0$ and $R=8a_0$ at the reference interaction, and the fixed-distance calculation uses $R=8a_0$ with $b$ ranges from $1$ to $5$. The internuclear-distance scan uses RABBITT sidebands; the interaction scan uses the continuous single-XUV-pulse wavepacket phase.

\subsection{Control of electron correlation and its effect on the Wigner time delay}
\label{sec:correlation_control}

The strength of the electron--electron interaction is controlled by the soft-core parameter \(b\) through
$
V_{ee}(x,x';b)=\frac{1}{\sqrt{(x-x')^2+b^2}}.
$
Increasing $b$ weakens the electron--electron repulsion while keeping the electron--nucleus interaction and internuclear distance unchanged. At fixed $R=8a_0$, varying $b$ therefore provides direct control over the electron--electron interaction strength and the degree of electronic correlation.

The electron--electron interaction induces configuration mixing in the two-electron wavefunction. The correlated ground state of \(\mathrm{H}_2\) can be expanded in terms of the $gg$ and $uu$ two-electron configurations as
\begin{equation}
\Psi_0(x,x';b)=c_{gg}(b)\phi_g(x)\phi_g(x')+c_{uu}(b)\phi_u(x)\phi_u(x'),
\label{eq:configuration_mixing}
\end{equation}
where $phi_g$ and $\phi_u$ denote the gerade and ungerade one-electron molecular orbitals, respectively, and $c_{gg}$ and $c_{uu}$ are their configuration coefficients. The electron--electron interaction couples these configurations and determines their relative weights in the ground-state wavefunction. As $b$ increases, the reduced electron--electron repulsion suppresses correlation-induced configuration mixing, decreasing the contribution of the $uu$ configuration relative to the $gg$ configuration. Thus, tuning $b$ changes the relative contributions of the two configurations to the correlated ground state and weakens the strength of electron correlation.

The redistribution of configuration weights directly affects the photoionization dynamics. The \(gg\) and \(uu\) configurations contribute different complex amplitudes to the outgoing photoelectron wavepacket. The $u$ ionic channel photoionization amplitude can be expressed as
\begin{equation}
M_u(E;b)
=c_{gg}(b)M_{gu}(E;b)+c_{uu}(b)M_{uu}(E;b),
\label{eq:ionization_amplitude}
\end{equation}
where $M_{gu}$ and $M_{uu}$ represent the photoionization amplitudes associated with the corresponding configurations. Changing their relative contributions modifies the interference between ionization pathways and consequently the energy dependence of the photoelectron phase.

Therefore, varying $b$ modifies the Wigner time delay through the combined effects of electron--electron interaction strength, configuration mixing, and interference between the $g$ and $uu$ ionization contributions.

\subsection{Hubbard model to H$_2$ system}
%
We consider the half-filled two-site Hubbard Hamiltonian
\begin{equation}
H=-t\sum_{\sigma}(c_{1\sigma}^{\dagger}c_{2\sigma}
+c_{2\sigma}^{\dagger}c_{1\sigma})
+U\sum_{i=1}^{2}n_{i\uparrow}n_{i\downarrow},
\label{eq：hubbard}
\end{equation}
where \(t\) and \(U\) denote the hopping amplitude and on-site Coulomb repulsion, respectively. Following the self-consistent Hubbard-III formulation \cite{Hubbard1964III}, we impose particle–hole symmetry and a paramagnetic half-filled state, with $n_\uparrow=n_\downarrow=1/2$ and the energy origin at \(U/2\). The resulting self-consistency relation is
\begin{equation}
3\Omega'(E)=E-\frac{U^2/4}{E-F(E)},
\label{eq:self_consistency}
\end{equation}
where \(F(E)\) is the inverse local propagator and \(\Omega'(E)\) is the scattering contribution.
Unlike the continuous density of states assumed in the original Hubbard-III analysis, the noninteracting two-site system has two discrete single-particle energies, \(\varepsilon_\pm=\pm t\), yielding
\begin{equation}
\rho_0(\varepsilon)=\frac12
\left[\delta(\varepsilon-t)+\delta(\varepsilon+t)\right].
\label{eq:pho_twosite}
\end{equation}
Substituting Eq.~\ref{eq:pho_twosite} into the local Green's function gives
\begin{equation}
2\pi G_{ii}(E)
=\int\frac{\rho_0(\varepsilon)\,d\varepsilon}
{F(E)-\varepsilon}
=\frac{F(E)}{F^2(E)-t^2}.
\label{eq:green_function}
\end{equation}
Using the Hubbard-III self-consistency condition,
\begin{equation}
\Omega'(E)=F(E)-\frac{1}{2\pi G_{ii}(E)}
=\frac{t^2}{F(E)},
\label{eq:omega_greenfunction}
\end{equation}
and substituting Eq.~\ref{eq:omega_greenfunction} into Eq.~\ref{eq:self_consistency}, we obtain
\begin{equation}
EF^2+
\left(\frac{U^2}{4}-E^2-3t^2\right)F
+3t^2E=0.
\label{eq:etf}
\end{equation}
The discriminant of this quadratic equation is
\begin{equation}
\begin{aligned}
\mathcal D(E)
&=\left(\frac{U^2}{4}-E^2-3t^2\right)^2
-12t^2E^2\\
&=\left[E^2-\left(\frac U2+\sqrt3t\right)^2\right]
\left[E^2-\left(\frac U2-\sqrt3t\right)^2\right].
\end{aligned}
\label{critical_condition}
\end{equation}
The inner branch points coalesce at the Fermi energy \(E=0\) when \(\mathcal D(0)=0\), giving
\begin{equation}
U_c^{\mathrm{HIII}}=2\sqrt3\,t
\approx3.464\,t.
\label{eq:criticalUc}
\end{equation}
This value defines a characteristic spectral scale within the Hubbard-III self-consistent approximation. For the finite two-site system, it represents a formal branch-point coalescence.

\subsection{Multi-reference coefficients during bond stretching}
Let $g$ and $u$ be the gerade and ungerade spatial orbitals of H$_2$. In the two-determinant singlet sector, the electronic state is
\begin{equation}
|\Psi_{\rm H_2}(R)\rangle=c_1(R)|gg\rangle+c_2(R)|uu\rangle,
\qquad c_1^2(R)+c_2^2(R)=1.
\label{eq:h2coeff}
\end{equation}
The phase of the $uu$ configuration is included in the determinant convention. Its coefficients interpolate between the united-atom and dissociation limits,
\begin{equation}
\lim_{R\to0}(c_1,c_2)=(1,0),\qquad
\lim_{R\to\infty}(c_1,c_2)=\left(\frac{1}{\sqrt2},\frac{1}{\sqrt2}\right).
\label{eq:h2limits}
\end{equation}
The growing $uu$-determinant weight, $c_2^2:0\to1/2$, describes an increasing multireference component. The equal-weight dissociation limit reflects strong static correlation as the bonding and antibonding configurations become degenerate. The Hubbard description of H$_2$ and the dependence of its effective interactions on interatomic distance have also been studied in Ref.~\cite{Chiappe2007}. The stretch from $R=4a_0$ to $R=8a_0$ increases this correlation and changes the channel-resolved molecular delay from a positive to a negative feature. The $R=8a_0$ soft-core interaction scan independently demonstrates the influence of electron repulsion on the sign.

\subsection{Channel projection and RABBITT at different separations}
For the nuclear-distance comparison in main Fig.~1, an XUV attosecond pulse train combined with a delayed infrared dressing pulse drives the two-electron TDSE. At the end of the propagation, the outgoing-electron channel wavefunction is obtained by projecting onto a normalized residual H$_2^+$ eigenstate $\varphi_c$. A schematic projection is
\begin{equation}
\chi_c(x,t_f)=\int dx'\,\varphi_c^*(x')\,\Psi(x,x',t_f),
\label{eq:proj}
\end{equation}
with exchange symmetrization and continuum separation treated according to the two-electron implementation. The analysis selects the ungerade state $c=u$. For the fitted sideband at energy $E$, the pump--probe delay dependence is described by
\begin{equation}
S_E(\Delta t)=A_E+B_E\cos\big[2\omega_{\mathrm{IR}}\Delta t-\phi_E\big].
\end{equation}
An inferred one-photon delay is then extracted by separating the adopted XUV contribution and continuum--continuum (CC) phase, schematically
\begin{equation}
\tau_u^{\mathrm{RABBITT}}(E)=\frac{\phi_E-\phi_{\mathrm{XUV}}(E)-\phi_{\mathrm{CC}}(E)}{2\omega_{\mathrm{IR}}},
\label{eq:rabbitt}
\end{equation}
with all phase and sign conventions chosen consistently. This procedure produces a sideband-derived estimate of a channel-resolved Wigner-type delay.

\subsection{Single-attosecond-pulse phase derivative at fixed $R$}
For the interaction scan in main Fig.~2, a \emph{single attosecond XUV pulse} ionizes H$_2$ at fixed $R=8a_0$. The channel projection yields a continuous outgoing photoelectron wavefunction. Its momentum representation may be written
\begin{equation}
\widetilde\chi_u(p,t_f)=\frac{1}{\sqrt{2\pi}}\int dx\,e^{-ipx}\chi_u(x,t_f),
\end{equation}
from which $dP_u/dE$ and the momentum-dependent phase are obtained. With $E=p^2/2$ (atomic units) and a consistently defined outgoing-energy branch, the displayed delay is an energy derivative of the \emph{single-pulse} complex continuum amplitude,
\begin{equation}
\tau_u^{\mathrm{1XUV}}(E)=\hbar\,\partial_E\arg A_u(E),
\label{eq:singlepulse}
\end{equation}
where $A_u(E)$ denotes the extracted channel amplitude with the propagation and pulse phase reference used by the TDSE analysis. The continuous-wavepacket phase contains the free-propagation factor $e^{-iEt_f/\hbar}$, treated with the phase reference used in the TDSE extraction. All values in the $b$ scan share the same pulse protocol and phase reference.

The results show negative delay features for $b\leq2.2$ and positive features for $b\geq2.4$ among the sampled values. Modifying $b$ changes the electron--electron interaction across both the interacting initial state and the photoelectron--residual-electron dynamics. These are two aspects of the interaction-driven many-electron effect. The scan thus demonstrates that electron--electron interaction controls the delay sign at fixed nuclear geometry through both configuration correlation and outgoing-electron scattering.

\subsection{definition of the internuclear distance for different vibrational states}

$\chi_v(R)$ denote the normalized reduced radial vibrational wavefunction of the $X$ state at $J=0$. The probability density for the internuclear coordinate is
\begin{equation}
 p_v(R)=|\chi_v(R)|^2,
 \qquad \int_0^\infty p_v(R)\,dR=1.
 \label{eq:sm-r-density}
\end{equation}
We define the representative distance assigned to each target state as the global mode of this density,
\begin{equation}
 R_v^{\mathrm{peak}}=
 \operatorname*{arg\,max}_{R>0}p_v(R).
 \label{eq:sm-r-peak}
\end{equation}
For the three states considered here, this global maximum lies in the outermost probability lobe, as shown in Fig.~\ref{fig:si-h2-stirap}(b). The wavefunctions were reconstructed continuously in the sine basis, and the density maxima were located within the node-separated lobes. Increasing the DVR grid from 800 to 1200 points changes the reported peak positions by less than $2\times10^{-7}$~Bohr for the fixed potential model.

Equation~\eqref{eq:sm-r-peak} defines a modal coordinate; it differs from the full expectation value $\langle R\rangle_v=\int R p_v(R)\,dR$ and from a conditional average over a selected probability region. It is used only to associate a representative coordinate with each vibrational state for comparison with fixed-$R$ calculations. Each prepared state retains the complete distribution $p_v(R)$; assigning $R_v^{\mathrm{peak}}$ does not constitute a vibrationally averaged RABBITT calculation. In particular, the three peak positions span $4.180$--$5.643$~Bohr and should not be identified with the fixed $R=8a_0$ geometry of the TDSE comparison. A quantitative test of the predicted delay reversal would require photoionization calculations that account for the nuclear distributions and the difference between the electronic models.

The numerical grid convergence above is conditional on the finite-basis Born--Oppenheimer potentials. Changing from aug-cc-pVDZ to aug-cc-pVTZ shifts the pump wavelength for this same intermediate by approximately $0.84$~nm and the three Stokes wavelengths by approximately $0.35$, $0.85$, and $1.47$~nm, respectively. Thus, these are candidate model wavelengths rather than experimental tuning frequencies. The estimates omit nonadiabatic, relativistic, and QED corrections, as well as AC Stark shifts and pulse-driven population dynamics. Experimental implementation requires suitable coherent VUV fields, sufficiently strong couplings, and adiabatic pulse parameters; no transfer efficiency is inferred here.

\subsection{Candidate STIRAP preparation of selected vibrational states}
\label{sec:sm-stirap-peak}

The following molecular-spectroscopy calculation provides a separate estimate of a possible vibrational-state preparation route. It uses three-dimensional Coulombic H$_2$ electronic potentials, whereas the fixed-$R$ TDSE calculations use a one-dimensional soft-core electronic Hamiltonian. The vibrational levels and preparation wavelengths below therefore provide an experimental reference rather than eigenstates or excitation frequencies of the soft-core TDSE model.

We consider a STIRAP scheme \cite{Bergmann2015} connecting the initial state $|i\rangle=|X\,{}^1\Sigma_g^+,v=0,J=0\rangle$ to $|f_v\rangle=|X\,{}^1\Sigma_g^+,v,J=0\rangle$, with $v=12,13,14$, through the common intermediate state $|e\rangle=|B\,{}^1\Sigma_u^+,v'=6,J'=1\rangle$. The pump and Stokes fields couple $|i\rangle\leftrightarrow|e\rangle$ and $|f_v\rangle\leftrightarrow|e\rangle$, respectively. Under two-photon resonance, their field-free frequencies satisfy
\begin{align}
 h\nu_{\mathrm P} &= E_e-E_i, &
 h\nu_{\mathrm S}^{(v)} &= E_e-E_{f_v},
 \label{eq:sm-stirap-frequencies}
 \end{align}
The common intermediate state allows the pump frequency to remain fixed while the Stokes frequency selects the target state.

\begin{figure}[tbp]
\centering
\includegraphics[width=\textwidth]{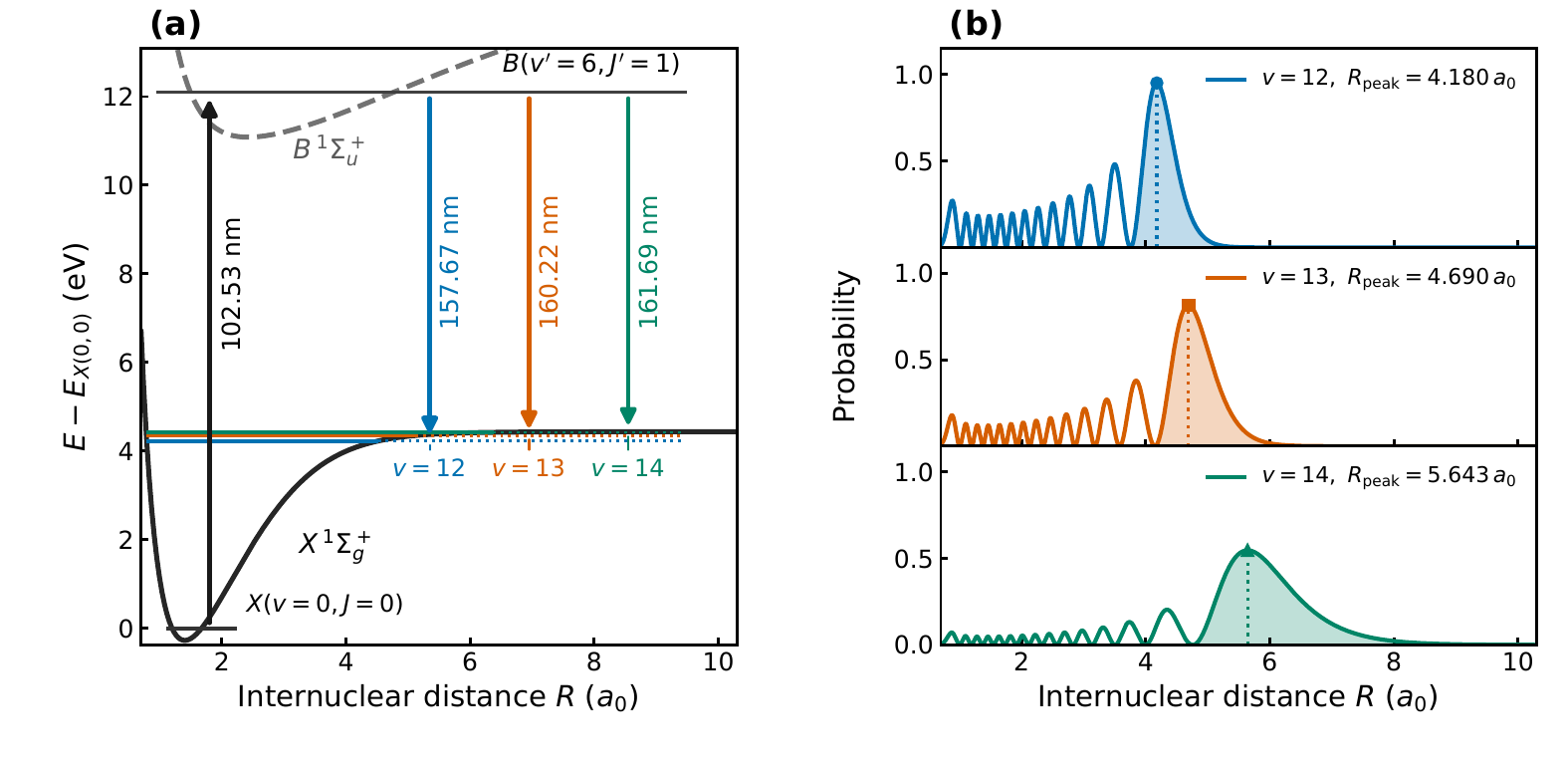}
\caption{Candidate STIRAP couplings and the density-peak definition of the representative internuclear distance. (a) FCI/aug-cc-pVTZ $X\,{}^1\Sigma_g^+$ (solid) and $B\,{}^1\Sigma_u^+$ (dashed) Born--Oppenheimer potential curves, referenced to $X(v=0,J=0)$. The black upward arrow denotes the pump coupling to $B(v'=6,J'=1)$ at $102.53$~nm. Colored downward arrows denote Stokes couplings to $X(v=12,13,14,J=0)$ at $157.67$, $160.22$, and $161.69$~nm, respectively. Horizontal level energies include the specified rotational contributions. Arrow positions along $R$ are schematic and indicate neither localized nuclear transitions nor temporal pulse order. (b) Normalized $J=0$ probability densities $p_v(R)$ for the three target states, overlaid on a single set of axes. Vertical dotted lines and markers identify the global density peaks at $R_v^{\mathrm{peak}}=4.180$, $4.690$, and $5.643$~Bohr. Colors consistently identify each target state across the two panels. The figure describes a candidate preparation pathway, not a simulated transfer efficiency.}
\label{fig:si-h2-stirap}
\end{figure}

The arrows in Fig.~\ref{fig:si-h2-stirap}(a) identify these field couplings and do not represent the temporal pulse order or transitions localized at the plotted horizontal coordinates. For STIRAP, the Stokes pulse precedes and overlaps the pump pulse. In the ideal adiabatic limit, the dark state evolves from $|i\rangle$ to $|f_v\rangle$ with suppressed intermediate-state population; the scheme does not require first populating the $B$ electronic state.

To estimate the field-free resonances and the target-state distributions, we calculated the $X\,{}^1\Sigma_g^+$ and $B\,{}^1\Sigma_u^+$ Born--Oppenheimer potential curves using FCI/aug-cc-pVTZ. Singlet electronic states were selected explicitly, using the $A_{1g}$ and $A_{1u}$ representations of $D_{\infty h}$, respectively. The electronic energies include nuclear repulsion and retain a common absolute energy zero. The electronic transition-dipole function $\mu_{XB}(R)$ was also calculated, with electronic phases made continuous between adjacent geometries using wavefunction overlaps.

The nuclear Hamiltonian for electronic state $\alpha=X,B$ was solved using the sine-DVR method,
\begin{equation}
 \hat H_{\alpha J}=
 -\frac{1}{2\mu}\frac{d^2}{dR^2}
 +V_\alpha(R)+\frac{J(J+1)}{2\mu R^2},
 \label{eq:sm-nuclear-hamiltonian}
\end{equation}
where atomic units are used and $\mu=m_p/2$ is the nuclear reduced mass. The potential grid spans $0.45$--$24$~Bohr, and natural cubic splines interpolate the electronic data. We used 800 DVR points, with 1200 points as a convergence check. The intermediate level includes the $J'=1$ rotational term. For the linearly polarized $J=0,M=0\leftrightarrow J'=1,M'=0$ path, the calculated pump transition-dipole magnitude is $|d_{\mathrm P}|=0.17490\,e a_0$; the Stokes magnitudes are $0.26547$, $0.44716$, and $0.18173\,e a_0$ for $v=12,13,14$, respectively. These values include the $1/\sqrt{3}$ rotational factor and support this level as a candidate shared intermediate. They do not establish a STIRAP transfer efficiency or a basis-converged optimal pathway.

\section{Small-lattice Hubbard models and photoelectron scattering}
\subsection{Periodic one- and two-dimensional geometries}
The small-lattice calculations employ a one-dimensional two-site periodic ring and a two-dimensional four-site periodic square described by the Hubbard model \cite{Hubbard1963}. In the site basis, the interacting many-electron Hamiltonian can be written
\begin{equation}
H_{\mathrm{Hub}}=-\sum_{ij\sigma} t_{ij}\,c^\dagger_{i\sigma}c_{j\sigma}
+U\sum_i n_{i\uparrow}n_{i\downarrow},\qquad t_{ij}=t_{ji}^*,
\label{eq:hubbard}
\end{equation}
where $t_{ij}$ includes the periodic connectivity. This matrix notation avoids double counting hopping bonds in a two-site ring. Diagonalization of the relevant $N$- and $(N-1)$-particle occupation-number Hamiltonians provides the interacting initial state $|\Psi_0^N\rangle$ and the residual-electron eigenstates $|\Phi_m^{N-1}\rangle$. The two-site and four-site geometries both implement periodic boundary conditions.

\subsection{Dyson orbitals as correlated emission sources}
The removal coefficients for site $j$ and spin $\sigma$ are
\begin{equation}
d_{mj}^{(\sigma)}=\langle\Phi_m^{N-1}|c_{j\sigma}|\Psi_0^N\rangle.
\label{eq:dysonSI}
\end{equation}
In terms of one-electron site orbitals $\varphi_j(r)$, the corresponding Dyson source can be represented as
\begin{equation}
\phi_{D,m}^{(\sigma)}(r)=\sum_j\varphi_j(r)\,d_{mj}^{(\sigma)}.
\label{eq:dysonorb}
\end{equation}
Thus the Hubbard eigenstates determine the correlated initial source and the remaining ionic state. The emitted photoelectron is subsequently propagated in a continuum scattering calculation. Dyson orbitals alone do not provide the final complex photoionization amplitude unless the outgoing-electron propagation is specified.

\subsection{Scattering from the residual electron and channel selection}
After the Dyson sources are constructed, the outgoing electron scatters from the residual electron occupying different ionic eigenstates. A schematic close-coupling form for the continuum channels is
\begin{equation}
\left[T_e+V_{\mathrm{nuc}}+E_c^{N-1}-E_{\mathrm{tot}}\right]\psi_c(r)
+\sum_m V_{cm}^{ee}(r)\psi_m(r)=S_c(r),
\label{eq:ccscattering}
\end{equation}
where $S_c$ is the source constructed from the Dyson orbital and the transition operator. The electron--electron scattering matrix element is
\begin{equation}
V_{cm}^{ee}(r)=\langle\Phi_c^{N-1}|V_{ee}(r,r_{\mathrm{bound}})|\Phi_m^{N-1}\rangle.
\end{equation}
This quantity describes the interaction of the outgoing photoelectron with the residual electron. The diagonal terms give channel-dependent scattering potentials, while off-diagonal terms may mix ionic channels. Equation~\eqref{eq:ccscattering} summarizes the coupled-channel scattering structure used to connect the Dyson source to outgoing amplitudes. The computed continuum and source amplitudes combine into the complex detected-channel amplitude
\begin{equation}
\M_c(E)=\sum_{m,j}K_{c,mj}(E)\,d_{mj},\qquad
\tau_c(E)=\hbar\,\operatorname{Im}\frac{\partial_E\M_c(E)}{\M_c(E)}.
\label{eq:kernelSI}
\end{equation}
Only contributions ending in the same coherent final ionic state $c$ interfere. Orthogonal residual ionic states that are not coherently selected must instead be summed as intensities, $I(E)=\sum_c|\M_c(E)|^2$. Changing site separation affects the correlated many-body source and the continuum-scattering matrix elements, generating the signed-delay features in main Fig.~3. Both lattice dimensions use periodic boundary conditions.

\subsection{Complex-amplitude direction and winding}
For $\M(E)=X(E)+iY(E)$, the local signed phase derivative is
\begin{equation}
\frac{\tau(E)}{\hbar}=\frac{X\partial_EY-Y\partial_EX}{X^2+Y^2}.
\label{eq:complexSI}
\end{equation}
The orientation of the energy-ordered trajectory determines the local angular phase change. For a continuous \emph{closed} nonzero-amplitude trajectory $C$, the integer winding number about the origin is
\begin{equation}
\nu=\frac{1}{2\pi}\oint_C d\arg\M
=\frac{1}{2\pi}\oint_C\frac{X\,dY-Y\,dX}{X^2+Y^2}\in\mathbb Z.
\label{eq:winding}
\end{equation}
The winding describes accumulated phase topology while the local angular direction determines the sign of $\tau(E)$; the connection between complex photoionization amplitudes, cross-section minima and delay sign is developed in Ref.~\cite{Ji2024}. Both are needed to characterize the full complex-amplitude phase structure; a winding number by itself does not fix the delay sign at every energy. For a nonclosed energy segment, the net phase change can be discussed but must not automatically be called an integer winding number. The complex-plane plot in main Fig.~4 supplies a geometrical representation of these effects.

\subsection{Key results from exact diagonalization and first-order photoionization theory}
For $N$ electrons, diagonalize the Hubbard Hamiltonian in the occupation-number (Hartree--Fock determinant) bases of the $N$- and $(N-1)$-particle sectors. With
\begin{align}
 |\Psi_0^{(N)}\rangle&=\sum_\alpha U_{0\alpha}^{(N)}|\phi_\alpha^{(N)}\rangle,
 &|\Psi_m^{(N-1)}\rangle&=\sum_\beta V_{m\beta}^{(N-1)}|\phi_\beta^{(N-1)}\rangle,
 \label{eq:HFexpandSI}
\end{align}
the removal matrix element is computed directly from the determinant overlaps,
\begin{equation}
 d_m^{(j\sigma)}=
 \langle\Psi_m^{(N-1)}|c_{j\sigma}|\Psi_0^{(N)}\rangle
 =\sum_{\alpha\beta}(V_{m\beta}^{(N-1)})^*U_{0\alpha}^{(N)}
 \langle\phi_\beta^{(N-1)}|c_{j\sigma}|\phi_\alpha^{(N)}\rangle.
 \label{eq:HFcoherentSI}
\end{equation}
The momentum-space annihilation operator $c_{k\sigma}=N_s^{-1}\sum_j e^{-ik r_j}c_{j\sigma}$ produces the corresponding Fourier-resolved amplitude. The key distinction is that the removal \emph{amplitude} sums the complex matrix elements before taking the phase, whereas the one-particle Green function uses the products of each matrix element and its conjugate.

In the first-order time-dependent derivation, the system and free-photoelectron Hamiltonians are separated as $H_0=H_{\rm Hub}+H_{\rm free}$ and the light interaction has the form
\begin{equation}
 H_{\rm int}(t)=\mathcal E(t)\hat\mu_+e^{-i\Omega_{\rm ph}t}+\mathrm{h.c.},
 \qquad \hat\mu_+=\sum_{qk\sigma}M_{qk}d_{q\sigma}^\dagger c_{k\sigma}.
 \label{eq:firstorderlightSI}
\end{equation}
Under the sudden dipole approximation $M_{qk}\simeq M_0\delta(q-k)$, first-order perturbation theory gives, for an initial $N$-electron eigenstate and an outgoing electron with energy $\omega$,
\begin{equation}
 \psi_{k\sigma}(\omega)\propto\sum_m
 \frac{\langle\Psi_m^{(N-1)}|c_{k\sigma}|\Psi_0^{(N)}\rangle}
 {\omega+E_m^{(N-1)}-E_0^{(N)}-\Omega_{\rm ph}+i\eta},
 \qquad \tau_{k\sigma}(\omega)=\partial_\omega\arg\psi_{k\sigma}(\omega).
 \label{eq:TDSEkeySI}
\end{equation}
The overall fermionic exchange sign $(-1)^{N-1}$ is independent of energy and drops out of the derivative. 

\section{Relation between the Wigner time delay and the winding number}
\label{sec:delay-winding}

The relation between the Wigner time delay and the topology of the complex photoionization amplitude provides a geometric interpretation of the delay sign reversal. As demonstrated by Ji \emph{et al.} \cite{Ji2024}, the phase evolution of a complex transition amplitude near an ionization minimum depends not only on its magnitude but also on whether its trajectory encloses the origin of the complex plane.

For a complex photoionization amplitude $\M(E)=X(E)+iY(E)=|\M(E)|e^{i\phi(E)}$, the Wigner delay is determined by the energy derivative of its phase,

\begin{equation}
\tau(E)=\hbar\frac{\partial\phi(E)}{\partial E}
=\hbar\frac{X(E)\partial_EY(E)-Y(E)\partial_EX(E)}
{X^2(E)+Y^2(E)}.
\label{eq:winding-delay-local}
\end{equation}

Geometrically, $\tau(E)$ measures the rate at which the complex transition amplitude rotates around the origin as the photoelectron energy increases. For a closed trajectory that does not cross the origin, the accumulated phase is characterized by the integer winding number,

\begin{equation}
\nu=\frac{1}{2\pi}\oint_C d\phi
=\frac{1}{2\pi\hbar}\oint_C\tau(E)\,dE,
\label{eq:winding-delay-integral}
\end{equation}

where the sign of $\nu$ specifies the orientation of the trajectory. The winding number remains unchanged under continuous deformations of the trajectory unless the amplitude passes through the origin, where its phase becomes undefined.

The connection between winding and the sign of the time delay is particularly apparent near destructive interference. Following the analysis of Ji \emph{et al.} \cite{Ji2024}, a photoionization amplitude near an antiresonance can be represented by a Fano-like expression,

\begin{equation}
D_s(\epsilon)=D_0
\frac{\epsilon+Q+is\gamma}{\epsilon+i},
\qquad
\epsilon=\frac{E-E_r}{\Gamma/2},
\qquad s=\pm1,
\label{eq:winding-fano}
\end{equation}

where $E_r$ and $\Gamma$ characterize the resonance energy and width, $Q$ describes the asymmetry, and $\gamma>0$ characterizes the displacement of the amplitude zero from the real-energy axis. The two choices of $s$ give identical photoionization intensities,

\begin{equation}
|D_s(\epsilon)|^2
=|D_0|^2
\frac{(\epsilon+Q)^2+\gamma^2}{\epsilon^2+1},
\label{eq:winding-intensity}
\end{equation}

but different phase derivatives,

\begin{equation}
\tau_s(E)=\frac{2\hbar}{\Gamma}
\left[
\frac{1}{\epsilon^2+1}
-\frac{s\gamma}{(\epsilon+Q)^2+\gamma^2}
\right].
\label{eq:winding-fano-delay}
\end{equation}

The interference-sensitive term changes sign with $s$, even though the photoionization intensity remains identical. In this model, the $s=+1$ trajectory has zero winding, whereas the $s=-1$ trajectory encloses the origin once. The corresponding narrow delay feature changes from negative to positive. Thus, the winding of the complex amplitude contains phase information that cannot be determined from the photoionization intensity alone.

To connect this geometric picture to the periodic Hubbard calculations, we consider a simple clockwise circular trajectory near destructive interference,

\begin{equation}
\M(E)=C+\rho e^{-i\theta(E)},
\qquad
C,\rho>0,\quad \partial_E\theta>0,
\label{eq:winding-circle}
\end{equation}

where $C$ represents a background contribution and $\rho$ describes the magnitude of an interfering contribution. The corresponding Wigner delay is

\begin{equation}
\tau(E)=
-\hbar
\frac{\rho[\rho+C\cos\theta(E)]}
{C^2+\rho^2+2C\rho\cos\theta(E)}
\frac{\partial\theta}{\partial E}.
\label{eq:winding-circle-delay}
\end{equation}

When $C>\rho$, the trajectory does not enclose the origin and its winding number is zero. When $\rho>C$, the origin lies inside the trajectory, giving a clockwise winding number of $\nu=-1$. At the point of closest approach to the origin, $\theta=\pi$, Eq.~\eqref{eq:winding-circle-delay} reduces to

\begin{equation}
\left.\tau(E)\right|_{\theta=\pi}
=
\hbar\frac{\rho}{C-\rho}
\frac{\partial\theta}{\partial E}.
\label{eq:winding-sign}
\end{equation}

The delay therefore changes from positive to negative as the trajectory evolves from $C>\rho$ to $\rho>C$. At the boundary $C=\rho$, the amplitude passes through the origin and its phase is undefined. This illustrates how a continuous change in the relative contributions of interfering pathways can induce both a change in winding number and a reversal of the Wigner delay near an amplitude minimum.

In the periodic Hubbard model, changing the intersite separation modifies the correlated initial state and the relative contributions of coherent electron-removal pathways. The resulting complex photoionization trajectories are shown in Fig.~4 of the main text. In the clockwise geometric picture, trajectories with zero winding exhibit a positive delay feature near the interference minimum, whereas trajectories enclosing the origin exhibit a negative feature. The calculated delay reversal is therefore associated with a change in the complex-amplitude geometry as electronic correlations evolve. The circular model provides a minimal illustration of this mechanism; the delays of the Hubbard system are obtained directly from the full complex photoionization amplitudes rather than from the circular approximation.

\section{Wigner time delay as an intensive thermodynamic quantity}
\label{sec:thermointensive}
The channel-resolved transition amplitude is a \emph{coherent} sum over removal eigenchannels. we use
\begin{equation}
 d_m^{(k\sigma)}=\langle\Psi_m^{(N-1)}|c_{k\sigma}|\Psi_0^{(N)}\rangle,
 \quad \omega_m=E_m^{(N-1)}-E_0^{(N)}-\Omega_{\rm ph},\quad
 D_{k\sigma}(\omega)=\sum_m\frac{d_m^{(k\sigma)}}{\omega+\omega_m+i\eta},
 \label{eq:thermoDfull}
\end{equation}
\begin{equation}
 \tau_{k\sigma}(\omega)=\frac{\partial}{\partial\omega}\arg D_{k\sigma}(\omega)
 =\operatorname{Im}\left[\frac{\partial_\omega D_{k\sigma}(\omega)}{D_{k\sigma}(\omega)}\right].
 \label{eq:thermodelay}
\end{equation}
In the lattice-model sections $\hbar=1$. The thermodynamic variable is the number of sites or momentum cells $L$. We examine the scaling of a real-valued size prefactor in the coherent amplitude.

\paragraph{Real-valued size prefactors.}
Suppose the leading system-size dependence of the complex amplitude has a real multiplicative factor, for example
\begin{equation}
 D(\omega;L)\propto L^\alpha[A(\omega)+iB(\omega)],
 \quad A,B\in\mathbb R,
 \label{eq:real-size}
\end{equation}
where $A$ and $B$ do not depend on $L$. The size factor contributes no phase. Therefore
\begin{equation}
\begin{split}
\arg D(\omega;L)&=\arg[A(\omega)+iB(\omega)],\\
\tau(\omega)&=\partial_\omega\tan^{-1}\frac{B(\omega)}{A(\omega)}=O(1)
\qquad(L\to\infty),
\end{split}
\label{eq:real-size-delay}
\end{equation}
with the inverse tangent understood with the continuous complex-phase branch. Thus, the size-dependent real normalization does not alter the Wigner phase derivative. The projected BCS analysis in Sec.~\ref{sec:bcs-full} likewise isolates the real amplitude prefactors from the phase contributing to the time delay.

The two quantities $D$ and $G$ are not interchangeable. The transition amplitude $D$ is the coherent sum of complex electron-removal channel amplitudes, whose relative phases are retained. By contrast, the Lehmann representation of the one-particle Green function $G$ contains products of each removal matrix element with its complex conjugate; its spectral weights therefore depend on squared matrix-element magnitudes. The interference terms between distinct channels that appear when taking the squared magnitude of a coherent sum are not separately encoded in these diagonal Lehmann weights. Consequently, an energy derivative of the phase of $G$, including one reconstructed from a spectral function by Hilbert transformation, is not generally equal to $\partial_\omega\arg D$ in a multichannel process. Throughout the remainder of this Supplementary Information, ``Wigner delay'' denotes Eq.~\eqref{eq:thermodelay}.

\section{Superconducting Hatsugai--Kohmoto model}
\subsection{Pairing mean-field Hamiltonian}
For a $d$-dimensional HK model \cite{Hatsugai1992}, with the superconducting pairing interaction treated as in Ref.~\cite{Phillips2020}, we write
\begin{equation}
H=\sum_k\left[\epsilon_k(n_{k\uparrow}+n_{k\downarrow})+U n_{k\uparrow}n_{k\downarrow}\right]-g\hat\Delta^\dagger\hat\Delta,
\qquad \hat\Delta=\sum_k c_{-k\downarrow}c_{k\uparrow},
\label{eq:hkbase}
\end{equation}
where $\epsilon_k=-(W/2d)\sum_{\nu=1}^d\cos k_\nu$, $\zeta_k=\epsilon_k-\mu$, and $g>0$ is the $s$-wave pairing strength. Following the mean-field decoupling, the Hamiltonian is block diagonal over $k>0$,
\begin{align}
H_{\rm MF}&=\sum_{k>0}H_k,\notag\\
H_k&=\zeta_k\sum_{\sigma}(n_{k\sigma}+n_{-k\sigma})
+U(n_{k\uparrow}n_{k\downarrow}+n_{-k\uparrow}n_{-k\downarrow})\notag\\
&\quad+\Delta^*(b_k+b_{-k})+\Delta(b_k^\dagger+b_{-k}^\dagger),\qquad
b_k^\dagger=c_{k\uparrow}^\dagger c_{-k\downarrow}^\dagger.
\label{eq:HKblock}
\end{align}
The pairing parameter $\Delta$ is the complex gap amplitude. In the weak-pairing regime, the self-consistent gap equation is
\begin{equation}
1=\frac g2\int d\xi\,\frac{N''(\xi)}{\sqrt{\xi^2+|\Delta|^2}},\qquad
\Delta\simeq\sqrt{(W-U)U}\,e^{-W/(2g)}.
\label{eq:HKgap}
\end{equation}
The half-filled chemical potential is $\mu=U/2$. The block has four fermionic modes $(k\uparrow,k\downarrow,-k\uparrow,-k\downarrow)$ and thus 16 occupation-number basis states. Diagonalization gives
\begin{equation}
H_k|\psi_n^{(k)}\rangle=E_n^{(k)}|\psi_n^{(k)}\rangle,
\quad |\Psi_{\mathbf n}\rangle=\mathcal A\!\bigotimes_{k>0}|\psi_{n_k}^{(k)}\rangle,
\quad E_{\mathbf n}=\sum_{k>0}E_{n_k}^{(k)},
\label{eq:HKeigen}
\end{equation}
where $\mathcal A$ ensures fermionic antisymmetrization.

\subsection{Zero-temperature Wigner delay}
At zero temperature only the many-body ground state appears on the ket side. Summing the final electronic states $\alpha$ over the complete Hilbert space gives
\begin{equation}
D_0^{(k\sigma)}(\omega)=\sum_\alpha
\frac{\langle\Psi_\alpha|c_{k\sigma}|\Psi_0\rangle}{\omega+E_\alpha-E_0-\Omega_{\rm ph}+i\eta}.
\label{eq:HKzeroFull}
\end{equation}
Because $c_{k\sigma}$ acts on a single $(k,-k)$ block, the other momentum sectors factorize and the expression reduces to the 16-state eigenbasis,
\begin{equation}
D_{0,k\sigma}(\omega)=\sum_{n=0}^{15}
\frac{\langle\psi_n^{(k)}|c_{k\sigma}|\psi_0^{(k)}\rangle}{\omega+E_n^{(k)}-E_0^{(k)}-\Omega_{\rm ph}+i\eta}.
\label{eq:HKzero}
\end{equation}
Define $a_{n0}=\langle\psi_n^{(k)}|c_{k\sigma}|\psi_0^{(k)}\rangle$ and $q_{n0}=\omega+E_n^{(k)}-E_0^{(k)}-\Omega_{\rm ph}+i\eta$. The phase derivative is computed analytically:
\begin{equation}
D'_{0,k\sigma}(\omega)=-\sum_n\frac{a_{n0}}{q_{n0}^2},\qquad
\tau_{0,k\sigma}(\omega)=\operatorname{Im}\frac{D'_{0,k\sigma}(\omega)}{D_{0,k\sigma}(\omega)}.
\label{eq:HKzeroDerivative}
\end{equation}

\subsection{Finite-temperature Wigner delay}
At finite temperature the initial ket is a thermal superposition of Hamiltonian eigenstates with amplitudes $e^{-\beta E_m/2}/\sqrt{Z}$, with $\beta=1/T$ and partition function $Z=\sum_m e^{-\beta E_m}$. The coherent transition amplitude becomes
\begin{equation}
D_T^{(k\sigma)}(\omega)=\frac{1}{\sqrt{Z}}
\sum_{\alpha,\alpha'}\frac{e^{-\beta E_{\alpha'}/2}
\langle\Psi_\alpha|c_{k\sigma}|\Psi_{\alpha'}\rangle}{\omega+E_\alpha-E_{\alpha'}-\Omega_{\rm ph}+i\eta}.
\label{eq:HKfiniteFull}
\end{equation}
The factor $e^{-\beta E_{\alpha'}/2}$ is the Boltzmann amplitude weight and $Z^{-1/2}$ normalizes the thermal ket. Reduction to the independent momentum block, gives
\begin{align}
D_{T,k\sigma}(\omega)&=Z_k^{-1/2}\sum_{n,m=0}^{15}
\frac{e^{-\beta E_m^{(k)}/2}\langle\psi_n^{(k)}|c_{k\sigma}|\psi_m^{(k)}\rangle}{\omega+E_n^{(k)}-E_m^{(k)}-\Omega_{\rm ph}+i\eta},\\
Z_k&=\sum_{m=0}^{15}e^{-\beta E_m^{(k)}}.
\label{eq:HKfinite}
\end{align}
Its derivative is
\begin{equation}
\tau_{T,k\sigma}(\omega)=\operatorname{Im}\frac{D'_{T,k\sigma}(\omega)}{D_{T,k\sigma}(\omega)},\qquad
D'_{T,k\sigma}(\omega)=-\frac{1}{\sqrt{Z_k}}\sum_{n,m}
\frac{e^{-\beta E_m^{(k)}/2}\langle\psi_n^{(k)}|c_{k\sigma}|\psi_m^{(k)}\rangle}{[\omega+E_n^{(k)}-E_m^{(k)}-\Omega_{\rm ph}+i\eta]^2}.
\label{eq:HKfiniteDerivative}
\end{equation}
Taking $\beta\to\infty$ selects the initial ground state and restores Eq.~\eqref{eq:HKzero}. Thermal occupation of higher states introduces transitions with additional signed amplitudes and energy denominators, which contribute coherently to the phase derivative.

\subsection{Exact diagonalization and numerical evaluation}
We implement the four fermion modes in binary occupation-number order $(k\uparrow,k\downarrow,-k\uparrow,-k\downarrow)$; their annihilation operators have matrix elements $\langle s\setminus i|c_i|s\rangle=(-1)^{\sum_{j<i}n_j}$ for an occupied mode $i$. The $16\times16$ Hermitian matrix $H_k$ is diagonalized to obtain $E_n^{(k)}$ and the signed amplitude matrix $C_{nm}=\langle\psi_n^{(k)}|c_{k\uparrow}|\psi_m^{(k)}\rangle$. The frequency-resolved amplitudes and their exact energy derivatives are evaluated using Eqs.~\eqref{eq:HKzero}--\eqref{eq:HKfiniteDerivative}. The same block-diagonalization procedure evaluates the zero-temperature and finite-temperature coherent amplitudes and their frequency derivatives directly from the eigenvalues and signed transition matrix elements.

\section{Wigner time delay in projected BCS system}
\label{sec:bcs-full}
\subsection{BCS Hamiltonian with $p$-wave pairing}

I will introduce another solvable model that has multi-reference wavefunction.
More details of BCS $p$-wave pairing (BCS-pw) model can be found in {\href{http://dx.doi.org/10.1103/PhysRevB.79.180501}{this work}}.
The original 2D BCS-pw Hamiltonian is defined as
\begin{equation}
\hat{H} = \sum_{\bm{k}} \frac{{\bm{k}}^2}{2 m} \hat{n}_{\bm{k}} - 
\frac{G}{4 m} \sum_{\bm{k} \ne \bm{k}^{\prime}} (k_x - i k_y)
(k_x^{\prime} + i k_{y}^{\prime})  \hat{c}_{\bm{k}}^{\dagger} 
\hat{c}_{-\bm{k}}^{\dagger} \hat{c}_{-\bm{k}^{\prime} } 
\hat{c}_{\bm{k}^{\prime}}
\end{equation}
with momentum ${\bm{k}}$, mass $m$, and a dimensionless coupling constant $G > 0$.
Following the original work, we consider spinless fermion in the Hamiltonian,
$\hat{n}_{\bm{k}} = \hat{c}_{\bm{k}}^{\dagger} \hat{c}_{\bm{k}}$, 
instead of previously studied spinful fermion.

\paragraph{Mean-field Hamiltonian}

We introduce the mean-field treatment of BCS-pw Hamiltonian,
\begin{equation}
\hat{H}_{\rm{MF}} = \sum_{\bm{k}} \zeta_{\bm{k}} \hat{n}_{\bm{k}}
- \frac{1}{4} \sum_{\bm{k}} \left[ \hat{\Delta} \left(k_x - i k_y \right)
\hat{b}_{\bm{k}}^{\dagger} + {\rm{h.c.}} \right]
\end{equation}
with $\zeta_{\bm{k}} = \bm{k}^2 / 2 m - \mu/2$, 
$\mu$ the chemical potential, 
${\rm{h.c.}}$ the Hermitian conjugate, and
\begin{equation}
\hat{\Delta} = \frac{G}{m} \sum_{\bm{k}} (k_x + i k_y) \, 
\bigl\langle \hat{c}_{-\bm{k}} \hat{c}_{\bm{k}} \bigr\rangle
\end{equation}
The creation operator in the second line is essentially a bosonic operator
(since it involves two fermionic operators), and only allows single occupation,
\begin{equation}
\hat{b}_{\bm{k}}^{\dagger} = \hat{c}_{\bm{k}}^{\dagger} 
\hat{c}_{-\bm{k}}^{\dagger} ~~~~~~~~~~~  
\hat{b}_{\bm{k}}^{\dagger} \hat{b}_{\bm{k}}^{\dagger} \ket{0} = 0
\end{equation}
The summation runs over ${\bm{\Omega}}^{+}$ where $k_{x} > 0$ and 
any $k_{y}$, so that we avoid the overcounting of energy levels.

\paragraph{Gap equation and phase diagram parameters}

The gap parameter $\Delta = | \hat{\Delta}| $ and chemical potential $\mu$ are determined from self-consistent gap equations,
\begin{subequations}
\begin{equation}
\sum_{\bm{k} \in \bm{\Omega}^{+}} \frac{\bm{k}^2}
{\sqrt{({\bm{k}}^2 - \mu)^2 + {\bm{k}}^2 \Delta^2}} = \frac{1}{G}
\end{equation}
\begin{equation}
\mu\sum_{\bm{k} \in \bm{\Omega}^{+}} \frac{1}
{\sqrt{({\bm{k}}^2 - \mu)^2 + {\bm{k}}^2 \Delta^2}} = 
2 M - L + \frac{1}{G}
\end{equation}
\end{subequations}
where we have set $m = 1$ for simplicity.
The filling fraction is given by $x = M / L$, for example $x=1/2$ for half filling.
with $L$ the number of sites (distinct ${\bm{k}}$ points in ${\bm{\Omega}}^{+}$), 
and $M$ the number of particles (Cooper pairs).
The ground state phase diagram is a 2D diagram of filling fraction $x$, 
and the inverse of rescaled coupling strength $1 / g = 1 / G L$.
More details of phase information can be found in Fig. 1 of reference work.

\paragraph{Bogoliubov transformation}

The mean-field Hamiltonian can be diagonalized by a Bogoliubov transformation.
For each ${\bm{k}} \in {\bm{\Omega}}^{+}$, we let
\begin{equation}
\begin{split}
\hat{c}_{\bm{k}} &= u_{\bm{k}} \hat{\gamma}_{\bm{k}} +  
v_{\bm{k}} \hat{\gamma}_{-\bm{k}}^{\dagger}  \\
\hat{c}_{-\bm{k}} &= u_{\bm{k}} \hat{\gamma}_{-\bm{k}}-  
v_{\bm{k}} \hat{\gamma}_{\bm{k}}^{\dagger}
\end{split}
\end{equation}
The diagonalized Hamiltonian reads
\begin{subequations}
\begin{equation}
\hat{H} = \sum_{\bm{k} \in {\bm{\Omega}}^{+}} E_{\bm{k}}
\left( \hat{\gamma}_{\bm{k}}^{\dagger}  \hat{\gamma}_{\bm{k}}
+ \hat{\gamma}_{-\bm{k}}^{\dagger}  \hat{\gamma}_{-\bm{k}} \right)
+ E_{0}
\end{equation}
\begin{equation}
E_{\bm{k}} = \frac{1}{2} \sqrt{({\bm{k}}^2 - \mu)^2 + 
{\bm{k}}^2 \Delta^2} ~~~~~~~~~~ E_{0} = \sum_{\bm{k} 
\in {\bm{\Omega}}^{+}} \left( \zeta_{\bm{k}} - E_{\bm{k}} \right)
\end{equation}
\end{subequations}
The complex-valued coefficients are given by 
\begin{equation}
|u_{\bm{k}}|^2 = \frac{E_{\bm{k}} + \zeta_{\bm{k}}}
{2 E_{\bm{k}}} ~~~~~~~~~~  |v_{\bm{k}}|^2 = 
\frac{E_{\bm{k}} - \zeta_{\bm{k}}}{2 E_{\bm{k}}} 
\end{equation}
with normalization $|u_{\bm{k}}|^2 + |v_{\bm{k}}|^2 = 1$ and relative phase
\begin{equation}
\frac{v_{\bm{k}}}{u_{\bm{k}}} = - \frac{d_{\bm{k}}}
{E_{\bm{k}} + \zeta_{\bm{k}}} ~~~~~~~~~~ d_{\bm{k}} = 
-\frac{\Delta}{2} \left( k_x - i k_y \right)
\end{equation}
For convention, we choose $u_{\bm{k}}$ to be real-valued, and write
\begin{equation}
v_{\bm{k}} = - e^{i \phi_{\bm{k}}} \sqrt{\frac{E_{\bm{k}} 
- \zeta_{\bm{k}}}{2 E_{\bm{k}}} } ~~~~~~~~~~
d_{\bm{k}} = \left| d_{\bm{k}} \right| \, e^{i \phi_{\bm{k}}}
\end{equation}
The momentum dependent phase factor $e^{i \phi_{\bm{k}}} 
\propto k_{x} - i k_{y}$ winds by $2\pi$ around ${\bm{k}} = 0$, 
which is the origin of the $p_{x} + i p_{y}$ topological structure.

\paragraph{Multi-reference ground state wavefunction}

The grand canonical ground state wavefunction can be expressed as 
\begin{equation}
\ket{{\rm{GS}}} = \ket{0_{\gamma}} = \prod_{\bm{k} \in \bm{\Omega}^{+}} 
\left[ u_{\bm{k}} + v_{\bm{k}} \hat{b}_{\bm{k}}^{\dagger} \right] \ket{0}
\end{equation}
However, here we want the ground state to have a fixed number of Cooper pairs.
After particle number projection, we have the $M$-pair ground state
\begin{equation}
\ket{\Psi_{M}} = \left[ \sum_{\bm{k} \in \bm{\Omega}^{+}} g_{\bm{k}} 
\,\hat{c}_{\bm{k}}^{\dagger}\hat{c}_{-\bm{k}}^{\dagger} \right]^{M} 
\ket{0} ~~~~~~~~~~ g_{\bm{k}} = \frac{2 E_{\bm{k}} - \bm{k}^2 + \mu}
{(k_x + i k_y) \Delta^{\ast}}  \propto \frac{v_{\bm{k}}}{u_{\bm{k}}}
\end{equation}
In thermodynamic limit, this state is equivalent to the particle-number projection
(up to unimportant normalization constant)
\begin{equation}
\ket{\Psi_{M}} = \mathcal{N}_{M} \int_{0}^{2 \pi } \frac{{\D}\theta}{2\pi} 
e^{-2  i M \theta} \exp\left( i \theta \hat{N} \right) \ket{0_\gamma} 
\end{equation}
where $\mathcal{N}_{M}$ is the normalization factor, and 
$\hat{N}$ is the total particle number operator,
\begin{equation}
\hat{N} = \sum_{\bm{k} \in \bm{\Omega}^{+}} \left( 
\hat{c}_{\bm{k}}^{\dagger} \hat{c}_{\bm{k}} + 
\hat{c}_{-\bm{k}}^{\dagger} \hat{c}_{-\bm{k}} \right)
\end{equation}
The energy of this state in the thermodynamic limit reads
(accounting for the constant term $E_{0}$ in mean-field Hamiltonian)
\begin{equation}
E_{M} = \frac{1}{2} \sum_{\bm{k} \in \bm{\Omega}{+}} 
{\bm{k}}^2 \left( 1 - \frac{2 \bm{k}^2 + \Delta^2 - 2 \mu}
{4 E_{\bm{k}}}\right) 
\end{equation}

\subsection{Transition amplitude}

Throughout this section, we only consider the transition amplitude using 
ground state $\ket{\Psi_{M}}$ as the initial state.
The excited state or finite temperature will be untouched, 
because they cannot be analytically (or even numerically) calculated.

\subsubsection{Finite size system at zero temperature}

\paragraph{Occupation numbering state of quasi-particles}

Before computing the transition amplitudes, we first specify 
the configuration state function (CSF), which will be used as 
the basis (eigen-basis) throughout this calculation.
We introduce the index vector ${\bm{n}} = (\cdots, n_{+\bm{k}}, n_{-\bm{k}} \cdots)$, where $n_{\pm\bm{k}} \in \{0, 1\}$ represents the occupation of Bogoliubov quasi-particle  $\hat{\gamma}_{\pm\bm{k}}$.
The CSF is defined as follow, which is also the eigenstate of mean-field Hamiltonian,
\begin{subequations}
\begin{equation}
\ket{\bm{n}} = \ket{\cdots, n_{+\bm{k}}, \, n_{-\bm{k}}, \, \cdots} = 
\prod_{\bm{k} \in  \bm{\Omega}^{+}}\left( \hat{\gamma}_{\bm{k}} 
\right)^{n_{+\bm{k}}} \left( \hat{\gamma}_{-\bm{k}}
\right)^{n_{-\bm{k}}} \ket{0_{\gamma}} 
\end{equation}
\begin{equation}
\hat{H} \ket{\bm{n}} = E_{\bm{n}} \ket{\bm{n}} ~~~~~~~~~~
E_{\bm{n}} = \sum_{\bm{k} \in \bm{\Omega}^{+}} E_{\bm{k}}
\left( n_{+\bm{k}} +  n_{-\bm{k}} \right) + E_{0}
\end{equation}
\end{subequations}
where $\ket{0_\gamma}$ is the vacuum state of quasi-particle 
(not to be confused with vacuum state of fermion $\ket{0}$).

\paragraph{$M$-pair ground state in thermodynamic limit}

The $M$-pair state $\ket{\Psi_M}$ serves as a linear combination of $\ket{\bm{n}}$ with fixed particle number $2 M$ (because of Cooper {\textit{pair}}).
Recall the integral form of $\ket{\Psi_M}$, and block diagonal form of numbering operator $\hat{N}$.
For each pair $({\bm{p}}, -{\bm{p}})$, we introduce the pair occupation state,
\begin{equation}
\ket{0, 0}_{\bm{p}} ~~~~~~~~~~ \ket{1, 1}_{\bm{p}} = 
\hat{\gamma}_{\bm{p}}^{\dagger} \hat{\gamma}_{-\bm{p}}^{\dagger} 
\ket{0_{\gamma}}
\end{equation}
The many-body state can be defined as the simple tensor product of these pair states,
since it involves even number of fermionic operators (0 or 2).
The action of $e^{i \theta \hat{N}}$ on $\ket{0_\gamma}$ then gives
\begin{equation}
e^{i \theta \hat{N}} \ket{0_{\gamma}} = \prod_{\bm{p} \in 
\bm{\Omega}^{+}}\left[ A_{\bm{p}}(\theta) \ket{0, 0}_{\bm{p}} 
+ B_{\bm{p}}(\theta) \ket{1, 1}_{\bm{p}} \right]
\end{equation}
where coefficients are defined by
\begin{equation}
A_{\bm{p}}(\theta) = |u_{\bm{p}}|^2 + |v_{\bm{p}}|^2 \, e^{2 i \theta} 
~~~~~~~~~~ B_{\bm{p}}(\theta) = u_{\bm{p}}v_{\bm{p}} (e^{2 i \theta} - 1)
\end{equation}
The $\ket{\Psi_M}$ then reads
\begin{equation}
\ket{\Psi_M} \propto \int_{0}^{2 \pi } \frac{{\D}\theta}{2\pi}
e^{-2 \, i \, M \, \theta} \prod_{\bm{p} \in \bm{\Omega}^{+}}
\left[ A_{\bm{p}}(\theta) \ket{0, 0}_{\bm{p}} 
+ B_{\bm{p}}(\theta) \ket{1, 1}_{\bm{p}} \right]
\end{equation}
Let $z = e^{2 i \theta}$. 
The integral in $\ket{\Psi_M}$ then becomes a contour integral,
\begin{equation}
\int_{0}^{2 \pi} \frac{{\D}\theta}{2\pi} e^{-2 i (M-1)\theta} \,
f(e^{2 i \theta}) = \oint \frac{\D z}{4 \pi i} z^{-M} \, f(z)
\end{equation}
Using this form, we can express the normalization factor as
\begin{equation}
\left( \mathcal{N}_{M} \right)^2 = \bra{0_\gamma} 
\mathcal{P}_{M} \ket{0_{\gamma}} =  \left[  \oint
\frac{\D z}{4 \pi i} z^{-M - 1} \prod_{\bm{k} \in \bm{\Omega}^{+}} 
\left( |u_{\bm{k}}|^2 + |v_{\bm{k}}|^2 \, z \right) \right]^{-1}
\end{equation}
with $\mathcal{P}_{M}$ the idempotent particle number projection ($\theta$-integral).

\paragraph{Transition amplitudes}

Let us consider the transition amplitude as follows
\begin{equation}
d_{\bm{k}}({\bm{n}}, \omega) = \frac{\bra{\bm{n}} \hat{c}_{\bm{k}}
\ket{\Psi_M}}{\omega + E_{\bm{n}} - E_{M} + i \eta }
\end{equation}
Substituted into the integrand, we obtain
\begin{equation}
\begin{split}
\bra{\bm{n}} \hat{c}_{\bm{k}} e^{i \theta \hat{N}} \ket{0_{\gamma}} 
&= \delta_{n_{+\bm{k}}, 0} \, \delta_{n_{-\bm{k}}, 1} \,
v_{\bm{k}} \, e^{2 i \theta} \\
&\times \prod_{\bm{p} \in \bm{\Omega}^{+}}^{ \bm{p} \ne \bm{k}}
\left[ \delta_{n_{+\bm{p}}, 0} \, \delta_{n_{-\bm{p}}, 0} \, 
A_{\bm{p}}(\theta) + \delta_{n_{+\bm{p}}, 1} \, 
\delta_{n_{-\bm{p}}, 1} \, B_{\bm{p}}(\theta) \right]
\end{split}
\end{equation}
Only $(n_{+\bm{k}}, n_{-\bm{k}}) = (0, 1)$, and 
$(n_{+\bm{p}}, n_{-\bm{p}}) = (0, 0)$ or $(1,1)$ has non-zero contributions.
Using the contour integral, we obtain
\begin{equation}
\bra{\bm{n}} \hat{c}_{\bm{k}} \ket{\Psi_M} = v_{\bm{k}}
\oint \frac{\D z}{4 \pi i} z^{-M} \prod_{\bm{p} \in \bm{\Omega}^{+}
}^{ \bm{p} \ne \bm{k}} \delta_{n_{+\bm{p}}, n_{-\bm{p}}} \left[ 
A_{\bm{p}}(z) (1 - n_{\bm{p}}) + B_{\bm{p}}(z) \, n_{\bm{p}} \right]
\end{equation}
The total transition amplitude should be the sum over all possible $\ket{\bm{n}}$,
\begin{equation}
D_{\bm{k}}(\omega) = \sum_{\bm{n}} \frac{\bra{\bm{n}} \hat{c}_{\bm{k}}
\ket{\Psi_M}}{\omega + E_{\bm{n}} - E_M + i \eta }
\end{equation}
where the summation runs over all the combinations below,
\begin{equation}
n_{+\bm{k}} = 0, ~~~~~ n_{-\bm{k}} = 1, ~~~~~
n_{+\bm{p}} = n_{-\bm{p}} \in \{ 0, 1 \}
\end{equation}
Using the Fourier transform,
\begin{equation}
\frac{1}{\omega + i \eta} = -i\int_{0}^{\infty} {\D}t \,
e^{i (\omega + i \eta) t}
\end{equation}
we rewrite the total transition amplitude as
\begin{equation}
\begin{split}
D_{\bm{k}}(\omega) &= -i \, v_{\bm{k}} \, \int_{0}^{\infty} {\D}t \,
e^{i (\omega + E_{0} + E_{\bm{k}} - E_{M} + i \eta) t} \\
&\times \oint \frac{\D z}{4 \pi i} z^{-M} \prod_{\bm{p} \in 
\bm{\Omega}^{+}}^{ \bm{p} \ne \bm{k}} \left[ A_{\bm{p}}(z) 
+ B_{\bm{p}}(z) \, e^{2 i E_{\bm{p}} t} \right]
\end{split}
\end{equation}
In the second line, we have factorized the summation and production using distributive law,
\begin{equation}
\sum_{\{ \bm{n}_{\bm{p}} \}} \prod_{\bm{p} \ne \bm{k}} 
\left[ \cdots \right] = \prod_{\bm{p} \ne \bm{k}} 
\sum_{n_{\bm{p}} = 0, 1} \left[ \cdots \right]
\end{equation}
The above integral form of $D_{\bm{k}}(\omega)$ is exact for finite $L$.

\paragraph{Wigner time delay as thermodynamic quantity}

We present a brief conclusion: Wigner time delay $\tau(\omega)$ 
is an intensive quantity that has a well-defined thermodynamic limit.
To prove, we denote $L$ as the size of system, and $D(\omega)$ as the transition amplitude.
As $L \to \infty$, we consider several types of scaling.
In the first case, if the leading term of $D(\omega)$ scales as a 
real-valued function against $L$, for example polynomial,
\begin{equation}
D(\omega) \propto L^{\alpha} \, \left[A(\omega) + i B(\omega) \right]
\end{equation}
where $A(\omega)$ and $B(\omega)$ are real-valued constants that are independent of $L$.
Since real-valued scaling factor $L^{\alpha}$ has no effect on phase factor,
the $\tau(\omega)$ would be intensive ($O(1)$ as $L \to \infty$), 
\begin{equation}
\tau(\omega) = \frac{\partial}{\partial \omega} {\rm{arg}} D(\omega) = 
\frac{\partial}{\partial \omega} \tan^{-1} \frac{B(\omega)}{A(\omega)}
\end{equation}
In the second case, if $D(\omega)$ scales as complex-valued exponential function,
(up to unimportant real-valued factors)
\begin{equation}
D(\omega) \propto e^{i \alpha L + i \, F(\omega)} ~~~~~~~~~~
{\rm{arg}} D(\omega) = \alpha \, L + F(\omega)
\end{equation}
As $L \to \infty$, we have
\begin{equation}
\tau(\omega) = \frac{\partial}{\partial \omega} {\rm{arg}} 
D(\omega) = \frac{\partial}{\partial \omega} F(\omega) \sim O(1)
\end{equation}

Even in a special case that $L$ couples with frequency $\omega$,
\begin{equation}
D(\omega) \propto e^{i \omega \, \alpha \, L + i \, F(\omega)}  ~~~~~~~~~~
{\rm{arg}} D(\omega) = \omega \, \alpha \, L + F(\omega)
\end{equation}
In this case, we need to consider the relation between 
$D(\omega)$ and one-particle Green's function $G(\omega)$.
The former involves coherent sum over all the eigenstates,
while the later involves incoherent sum,
\begin{equation}
D(\omega) = \sum_{n} d_{n}(\omega) ~~~~~~~~~~
G(\omega) = \sum_{n} \left|d_{n}(\omega)\right|^2
\end{equation}
To have non-vanishing $e^{i \omega \alpha L}$ in transition amplitude,
but leave Green's function intensive as $O(1)$, we should have
\begin{equation}
D(\omega) = \sum_{n} e^{i \omega \alpha_n L} \, \bar{d}_{n}(\omega)
\end{equation}
On the other hand, according to typical photo-ionization experiment, 
each $d_{n}(\omega)$ is the expectation value of annihilation operators 
(or their linear summation form),
\begin{equation}
d_{n}(\omega) = \frac{\bra{n} \hat{c}_{k} \ket{\psi_0}}
{\omega + E_{n} - E_{0} + i \eta}
\end{equation}
The initial state $\ket{\psi_0}$ or eigenstate $\ket{n}$ may involve 
some scaling factors against $L$, but $\omega$ cannot be coupled with 
$L$ in an exponential form.
Thus, this special case does not exist in physical system, 
at least not in the theoretical framework of photo-ionization experiment.
Physically, this means that energy of incident photon $\omega$ has no relations 
with targeting system size $L$, which obeys our intuition.

\subsubsection{Generalization to thermodynamic limit}

In the following we consider the transition amplitude $D_{\bm{k}}(\omega)$ 
at thermodynamic limit, where $M \to \infty$, $L \to \infty$, but 
$x = M/L$ remains constant.

\paragraph{One-particle spectrum at thermodynamic limit}

In the thermodynamic limit, the sum over momentum is replaced by the integral over single-particle energy, $\varepsilon = k^2/2$.
For 2D system, the density of states per site in the half Brillouin zone ${\bm{\Omega}}^{+}$ is constant,
\begin{equation}
\rho(\varepsilon) = \frac{1}{\Omega_c} ~~~~~~~~~~ 
\varepsilon \in (0, \Omega_c)
\end{equation}
with $\Omega_{c}$ the ultraviolet cutoff.
Next, we introduce
\begin{equation}
\zeta_{\bm{k}} \to \zeta(\varepsilon) = \varepsilon - \frac{\mu}{2} 
~~~~~~~~~~ E_{\bm{k}} \to E(\varepsilon) = 
\frac{\sqrt{(2 \varepsilon - \mu)^2 + 2 \varepsilon \Delta^2}}{2}
\end{equation}
The energy-dependent BCS coefficients become
\begin{equation}
u_{\bm{k}} \to u(\varepsilon) = \sqrt{\frac{E(\varepsilon) + 
\zeta(\varepsilon) }{2 E(\varepsilon)}} ~~~~~~~~~~
v_{\bm{k}} \to e^{i \phi} \, v(\varepsilon) ~~~~~~~~~~ 
v(\varepsilon) = -\sqrt{\frac{E(\varepsilon) - 
\zeta(\varepsilon) }{2 E(\varepsilon)}},
\end{equation}

\paragraph{Saddle point equations for $z$ and $t$}

At thermodynamic limit, we can replace the summation in $D_{\bm{k}}(\omega)$ by the integral,
\begin{equation}
\begin{gathered}
\prod_{\bm{p} \ne \bm{k}} \left[ \cdots \right] = \exp\left[ 
\sum_{\bm{p} \ne \bm{k}} \left( \cdots \right) \right] \\
\sum_{\bm{p} \in \bm{\Omega}^{+}} f(E_{\bm{p}}) ~~~ \to ~~~
L \int_{0}^{\Omega_c} {\D}\varepsilon \, \rho(\varepsilon) 
\int_{-\pi/2}^{\pi/2} \frac{\D \phi}{\pi} f(E(\varepsilon))
\end{gathered}
\end{equation}
We further define the grand canonical potential per site,
\begin{subequations}
\begin{equation}
\Phi(z, t) = \int_{0}^{\Omega_c} {\D}\varepsilon \, \rho(\varepsilon) \, 
\int_{-\pi/2}^{\pi/2} \frac{\D \phi}{\pi} \ln \left[ A(\varepsilon, z) + 
e^{2 i E(\varepsilon) t} \, e^{i \phi} \, B(\varepsilon, z)  \right],
\end{equation}
\begin{equation}
A(\varepsilon, z)  = |u(\varepsilon)|^2 + |v(\varepsilon)|^2 \, z ~~~~~~~~~~ 
B(\varepsilon, z) = u(\varepsilon) \, v(\varepsilon) \, (z - 1)
\end{equation}
\end{subequations}
Up to $1/L$ contribution from ${\bm{p}} = {\bm{k}}$,
which is negligible in thermodynamic limit, we have
\begin{equation}
\prod_{\bm{p} \in \bm{\Omega}^{+}}^{ \bm{p} \ne \bm{k}} 
\left[ A_{\bm{p}}(z) + B_{\bm{p}}(z) \, e^{2 i E_{\bm{p}} t} 
\right] \simeq e^{L \, \Phi(z, t)}  
\end{equation}
The coefficient becomes a contour integral,
\begin{equation}
\oint \frac{\D z}{4 \pi i} z^{-M} \prod_{\bm{p} \in 
\bm{\Omega}^{+}}^{ \bm{p} \ne \bm{k}} \left[ A_{\bm{p}}(z) 
+ B_{\bm{p}}(z) \, e^{2 i E_{\bm{p}} t} \right] \simeq
\oint \frac{\D z}{4 \pi i} e^{L \, \left[ \Phi(z, t) - x \ln z \right]} .
\end{equation}
As $L \to \infty$, the contour integral is mainly determined by 
the saddle point equation (converting to Gaussian integral), 
\begin{equation}
\frac{\partial}{\partial z} F(z, t) = \frac{\partial}{\partial z}
\left[ L \, \Phi(z, t) - M \, \ln z \right] = 0
\end{equation}
After some algebra, we obtain the $z$-saddle-point equation for numbering conservation,
\begin{equation}
\int_{0}^{\Omega_c} {\D}\varepsilon \, \rho(\varepsilon) \, 
\int_{-\pi/2}^{\pi/2} \frac{\D \phi}{\pi}
\frac{z |v(\varepsilon)|^2 + e^{2 i E(\varepsilon) t} \, e^{i \phi} \, 
z \, u(\varepsilon) \, v(\varepsilon)}{A(\varepsilon, z) + 
e^{2 i E(\varepsilon) t} \, e^{i \phi} \, B(\varepsilon, z)} = x
\end{equation}
Denote $z_{0}(t)$ as the solution of this integral equation at time $t$.
Around saddle point $z_{0}(t)$, we write $z(t) = z_{0}(t) e^{i \theta}$ 
so that $\ln z(t) = \ln z_0(t) + i \theta$, with $\theta$ a dummy integration variable. 
The integrand is expanded by Taylor series up to second-order,
\begin{equation}
\begin{split}
F(z, t) &= \Phi(z, t) - x \, \ln z \\
&= F(z_0(t), t) + \frac{1}{2} \frac{\partial F}{\partial (\ln z)^2} 
\biggr|_{z_0(t)} \left( i \theta \right)^2 + \cdots
\end{split}
\end{equation}
The first-order derivative vanishes at the saddle point, 
while the second-order derivative reads
\begin{equation}
\sigma^2(z_0(t), t)  = - \frac{\partial^2 F(z, t)}{\partial (\ln z)^2}\biggr|_{z_0(t), t} 
\end{equation}
The contour integral is approximated by a Gaussian integral in $\theta$,
\begin{equation}
\begin{split}
\oint \frac{\D z}{4 \pi i} e^{L \, F(z,t)} &\simeq e^{L \, F(z_0(t), t)} 
\int {\D}\theta \, e^{- L \sigma_{z}^2(z_0(t), t) \, \theta^2} \\
&= \frac{e^{L \, [\Phi(z_0(t), t) - x \ln z_0(t)]}}{\sqrt{2 \pi L \sigma^2(z_0(t), t)}} 
\end{split}
\end{equation}

Inserting into the expression of transition amplitude, we obtain the $t$-integral form,
\begin{equation}
D_{\bm{k}}(\omega) \simeq -i \, \mathcal{N}_{M} \, v_{\bm{k}} \, 
\int_{0}^{\infty} {\D}t \frac{e^{\mathcal{S}(t)}}
{\sqrt{2 \pi L \sigma^2(z_0(t), t)}} 
\end{equation}
where the effective action is defined by
\begin{equation}
\mathcal{S}(t) = i (\omega + E_{0} + E_{\bm{k}} - E_{M}) t + 
L \left[ \Phi(z_0(t), t) - x \, \ln z_{0}(t) \right].
\end{equation}
This Fourier integral is again determined by the saddle point, 
$\mathcal{S}^{\prime}(t) = 0$.
By the envelope theorem, since $\partial_z [ \Phi(z, t) - x \ln z ] = 0$, we have
\begin{equation}
\frac{\D}{\D t} \left[ \Phi(z_0(t), t) - x \ln z_0(t) \right] = 
\frac{\partial \Phi}{\partial t}\biggr|_{z_0(t), t}
\end{equation}
Therefore, saddle point becomes
\begin{equation}
i \frac{\omega + E_{0} + E_{\bm{k}} - E_{M}}{L} + 
\frac{\partial \Phi}{\partial t} \biggr|_{z_0(t_0), t_0} = 0
\end{equation}
Written explicitly, we obtain the $t$-saddle point equation for energy conservation,
\begin{equation}
\frac{\omega + E_{0} + E_{\bm{k}} - E_{M}}{L} = 
- \int_{0}^{\Omega_c} {\D}\varepsilon \, \rho(\varepsilon) \int_{-\pi/2}^{\pi/2} 
\frac{\D \phi}{\pi}\frac{2 \,  E(\varepsilon) \, e^{2 i E(\varepsilon) t} 
\, e^{i \phi} \, B(\varepsilon, z_0) }{A(\varepsilon, z_0) + 
e^{2 i E(\varepsilon) t} \, e^{i \phi} \, B(\varepsilon, z_0)} 
\end{equation}
The left-hand side is further regularized by gap equation at $L \to \infty$,
\begin{equation}
\begin{split}
\mathcal{E} &= \frac{\omega + E_{0} + E_{\bm{k}} - E_{M}}{L} \\
&\xrightarrow{L \to \infty} \int_{0}^{\Omega_c} {\D}\varepsilon \, 
\rho(\varepsilon) \left[ - \frac{\mu}{2} - E(\varepsilon) + \frac{\varepsilon
\left( 4 \varepsilon + \Delta^2 - 2 \mu \right)}{4 E(\varepsilon)} \right]
+ \frac{\omega + E_{\bm{k}}}{L} \\
&\simeq -\frac{\mu}{2} - \frac{\Delta^2}{4 g} + \frac{\mu}{2}
\int_{0}^{\Omega_c} {\D}\varepsilon \, \rho(\varepsilon)
\frac{2 \varepsilon - \mu}{2 E(\varepsilon)} + O(1/L)
\end{split}
\end{equation}
If necessary, the normalization factor can be solved from another saddle point equation.
However, we notice that as a real-valued factor, $\mathcal{N}_{M}$ 
does not affect the phase of transition amplitude (see below).
\begin{subequations}
\begin{equation}
\prod_{\bm{k} \in \bm{\Omega}^{+}} \left( |u_{\bm{k}}|^2 
+ y \, |v_{\bm{k}}|^2 \right) \to \exp\left[ L \int_{0}^{\Omega_c} 
{\D}\varepsilon \rho(\varepsilon) \,\ln\left( \left| u(\varepsilon) \right|^2+  
y \, \left| v(\varepsilon) \right|^2  \right) \right]
\end{equation}
\begin{equation}
y \frac{\partial (\cdots)}{\partial y} = \int_{0}^{\Omega_c} 
{\D}\varepsilon \rho(\varepsilon) \frac{y \, |v(\varepsilon)|^2}
{|u(\varepsilon)|^2+ y \, |v(\varepsilon)|^2} = x
\end{equation}
\begin{equation}
\mathcal{N}_{M}  = \left[ \frac{e^{L \, \chi(y_0)} }
{\sqrt{2 \pi L \varsigma^2}}  \right]^{-1/2} ~~~~~~~~~~
\varsigma^2 = \frac{\partial^2 \chi(y)}{\partial y^2}\biggr|_{y_0}
\end{equation}
\begin{equation}
\chi(y) = \int_{0}^{\Omega_c} {\D}\varepsilon \rho(\varepsilon) \,
\ln\left( \left| u(\varepsilon) \right|^2+  y \, 
\left| v(\varepsilon) \right|^2  \right) - x \ln y
\end{equation}
\end{subequations}

\paragraph{Final expression}

Assume that the solution of three saddle point equations at given 
$\omega$ are $t_{0}$, $z_{0}$, and $y_{0}$.
The final expression of transition amplitude reads
\begin{equation}
\begin{split}
D_{\bm{k}}(\omega) &= -i \, v_{\bm{k}} \, \frac{e^{\mathcal{S}(t_0)}
\, e^{- L \chi(y_0)/2} \left( 2 \pi L \varsigma^2 \right)^{1/4}}
{\sqrt{2 \pi L \sigma^2(z_0, t_0)}\sqrt{-2 \pi L 
\mathcal{S}^{\prime\prime}(t_0)}} \\
&= -i \, v_{\bm{k}} \, e^{\mathcal{S}(t_0)} \, e^{i \theta(t_0)}
\, \mathcal{A}(z_0, t_0, y_0) \, O\left( L^{-3/4} \, e^{-L} \right)
\end{split}
\end{equation}
with $\mathcal{A}(z_0, t_0, y_0) $ a real-valued factor, and 
$\theta(t_0)$ a phase factor of Hessian of effective action,
\begin{equation}
\theta(t_0) = -\frac{1}{2} {\rm{arg}} \left[ -\mathcal{S}^{\prime\prime}(t_0) \right]
\end{equation}
As $L \to \infty$, $D_{\bm{k}}(\omega) \to 0$ vanishes.
In conclusion, $D_{\bm{k}}(\omega)$ itself is not a well-defined quantity in thermodynamic limit.

\subsection{Wigner time delay at thermodynamic limit}

\subsubsection{Analytical derivations from transition amplitude}

\paragraph{Wigner time delay from transition amplitude}

Although $D_{\bm{k}}(\omega)$ vanishes as $L \to \infty$, 
its phase factor remains finite since $L^{-3/4}$ and 
$\mathcal{A}(z_0, t_0, u_0)$ are real-valued factors,
\begin{equation}
{\rm{arg}} D_{\bm{k}}(\omega) = {\rm{Im}} \mathcal{S}(t_0) + 
\theta(t_0) + {\rm{const}}
\end{equation}
where the last constant factor comes from the phase of 
$(-i \, v_{\bm{k}})$, which is $\omega$-independent.
Because both $z_{0} \sim z_{0}(t_0)$ and $t_{0} \sim t_0(\omega)$ are saddle points, we have for envelope theorem,
\begin{equation}
\frac{\D}{\D \omega} \mathcal{S}(t_0) = \frac{\partial}{\partial \omega} 
\mathcal{S}(t_0) = i \, t_{0}(\omega) ~~~~~~~~~~
\frac{\partial}{\partial \omega} {\rm{Im}} 
\mathcal{S}(t_0)  = {\rm{Re}} \, t_{0}(\omega)
\end{equation}
Regarding the second term $\theta(t_0)$, we expand near a critical point using Taylor series, 
\begin{equation}
\begin{split}
\mathcal{S}^{\prime\prime}(t_0) &= L \, f^{\prime\prime} (t_0) ~~~~~~~~~~
f(t_0) = \Phi(z_0, t_0) - x \, \ln z_0 \\
&\simeq f^{\prime\prime}(t_c) + f^{\prime\prime\prime}(t_c) \, (t_0 - t_c) + \cdots\\ 
&\simeq L \, c \, (t_0 - t_c)
\end{split}
\end{equation}
with $t_{c}$ the critical value, and $c = f^{\prime\prime\prime}(t_c)$. 
Then, we have
\begin{equation}
\theta(t_0) = -\frac{1}{2} {\rm{arg}}\left[ - L c (t_0 - t_c) \right] 
\end{equation}
Taking the full derivative,
\begin{equation}
\begin{gathered}
i + L \, f^{\prime\prime}(t_0) \, \frac{\D t_0}{\D \omega} = 0 \\
\frac{\D}{\D \omega}\theta(t_0) = -\frac{1}{2} \, 
{\rm{Im}} \left[ \frac{1}{t_0 - t_c} \frac{\D t_0}{\D \omega} \right]
\end{gathered}
\end{equation}
From the saddle point equation, we know that
\begin{equation}
\mathcal{S}^{\prime}(t_0) = i \, \Omega + L \, f^{\prime}(t_0) = 0
~~~~~~~~~~ \Omega = \omega + E_{0} + E_{\bm{k}} - E_{M}
\end{equation}
\begin{equation}
\frac{\D t_0}{\D \omega} = -\frac{i}{L f^{\prime\prime}(t_0)} 
\simeq -\frac{i}{L \, c \, (t_0 - t_c)}
\end{equation}
After substitution, we obtain
\begin{equation}
\frac{\D}{\D \omega}\theta(t_0) = \frac{1}{2 \, L} 
{\rm{Re}} \left[ \frac{1}{c (t_0 - t_c)^2}\right] 
\end{equation}
Collecting all the terms, we have for Wigner time delay,
\begin{equation}
\tau_{\bm{k}}(\omega) = \frac{\D}{\D \omega} {\rm{arg}} 
D_{\bm{k}}(\omega) = {\rm{Re}} \left[ t_0(\omega) \right]
+ \frac{1}{2 \, L} {\rm{Re}} \left[ \frac{1}{c (t_0 - t_c)^2}\right] 
\end{equation}

Let us look more carefully for $t_0 \to t_c$.
Near critical point, we expand $\delta\omega = \omega - \omega_c$.
At $\omega = \omega_c$, we have
\begin{equation}
\mathcal{S}^{\prime}(t_c) = i (\Omega_c + \delta \omega) 
+ L \left[ f^{\prime}(t_c) + \frac{f^{\prime\prime\prime}(t_c) }
{2}(t_0 - t_c)^2 + \cdots\right] = 0
\end{equation}
The constant term cancels out, and up to leading order we have
\begin{equation}
(t_0 - t_c)^2 \simeq -\frac{2 i \delta \omega}{L f^{\prime\prime\prime}(t_c)}
\end{equation}
Substituted into the time delay, we can eliminate the $L$-factor,
\begin{equation}
\frac{\D}{\D \omega}\theta(t_0) \simeq  \frac{1}{\delta \omega} 
\, {\rm{Re}} \left[  \frac{i \, f^{\prime\prime\prime}(t_c)}{4 c} \right]
\sim O \left( \frac{1}{\delta \omega} \right) =  \frac{1}{\delta \omega} 
\, {\rm{Re}}\left[  \frac{i}{4} \right] = 0
\end{equation}
The last equality arises from our assumption $f^{\prime\prime\prime}(t_c) = c$.
Thus, we conclude that time delay is always a smooth function,
\begin{equation}
\tau_{\bm{k}}(\omega) = \frac{\D}{\D \omega} {\rm{arg}} 
D_{\bm{k}}(\omega) = {\rm{Re}} \left[ t_0(\omega) \right]
\end{equation}

\paragraph{Saddle point evaluation of integral}

Recall that in our previous derivations, all the integration are carried out using saddle point technique.
This treatment breaks down when second-order derivative vanishes,
that is $\mathcal{S}^{\prime\prime}(t_0) \to 0$.
To see, let us write 
\begin{equation}
\mathcal{S}(t) = i \Omega t + L \, g(z, t), ~~~~~~~~~~
g(z, t) = \Phi(z, t) - x \ln z
\end{equation}
We also denote $f^{(z)}= {\partial} f / {\partial}z$, and 
$f^{(t)} = {\partial} f / {\partial}t$.
The saddle points of $z$ and $t$ are (see previous subsection)
\begin{equation}
g^{(z)}(z_0, t) = 0 ~~~~~~~~~~ \mathcal{S}^{(t)}(t_0) 
= i \Omega + L \, g^{(t)}(z_0, t) = 0
\end{equation}
The second-order derivative at saddle points is
\begin{equation}
\mathcal{S}^{\prime\prime}(t_0)  = L \, f^{\prime\prime}(t_0) 
~~~~~~~~~~ f(t) = g(z_0(t), t)
\end{equation}
Using the chain rule, $-z_{0}^{\prime}(t) = g^{(t z)} / g^{(z z)}$, we have
\begin{equation}
f^{\prime\prime}(t_0) = \frac{{\rm{det}} H(z_0(t_0), t_0)}
{g^{(z z)}(z_0(t_0), t_0)} ~~~~~~~~~~ H(z, t) = 
\begin{pmatrix}
g^{(t t)}(z, t)  & g^{(t z)}(z, t)  \\ g^{(t z)}(z, t)  & g^{(z z)}(z, t) 
\end{pmatrix}
\end{equation}
Recall the explicit expression of functions $f$ and $g$. 
We have for various derivatives
\begin{subequations}
\begin{equation}
g^{(z z)}(z, t)  = \int {\D}\varepsilon \rho(\varepsilon) \int \frac{\D \phi}
{\pi} \left[ - \frac{(\partial_z W)^2}{W^2} \right] + \frac{x}{z^2} 
\end{equation}
\begin{equation}
g^{(t t)}(z, t)  = \int {\D}\varepsilon \rho(\varepsilon) \int 
\frac{\D \phi}{\pi} \left[ \frac{\partial_t^2 W}{W} - 
\frac{(\partial_t W)^2}{W^2} \right]  
\end{equation}
\begin{equation}
g^{(t z)}(z, t)  = \int {\D}\varepsilon \rho(\varepsilon) \int 
\frac{\D \phi}{\pi} \left[ \frac{\partial_t \partial_z W}{W} - 
\frac{(\partial_t W)(\partial_z W)}{W^2} \right]
\end{equation}
\begin{equation}
W(z, t, \varepsilon, \phi) = A(\varepsilon, z) + e^{2 i E(\varepsilon) t} 
\, e^{i \phi} \, B(\varepsilon, z) 
\end{equation}
\end{subequations}
The saddle point treatment fails and critical point exists when 
$\mathcal{S}^{\prime\prime}(t_0) \propto {\rm{det}} H(z_0(t_0), t_0) = 0$.

\subsection{Time delay near phase boundary lines}

Near phase boundary lines, there are two features of Wigner time delay $\tau_{\bm{k}}(\omega)$ when we move across the phase boundary lines:
\begin{itemize}
\item Singularity of derivative $\partial_\xi \tau_{\bm{k}}(\omega, \xi)$;
\item Discontinuous jump of $\tau_{\bm{k}}(\omega, \xi)$;
\end{itemize}
Here, we have assumed fixed frequency $\omega$ and momentum ${\bm{k}}$.
The $\xi$ is the coordinate along a path in phase diagram (see below),
which will move across the Moore-Read $(\mu = \Delta^2/4)$ or Read-Green $(\mu = 0)$ lines.

\subsubsection{Analytical derivations}

\paragraph{Path $\xi$ along phase diagram}

Define $\xi$ as a real coordinate along a path in the phase diagram:
\begin{equation}
x=x(\xi),\qquad g=g(\xi).
\end{equation}
Let $\xi_c$ be the value at which the path crosses a boundary.
The functions $\mu(\xi)$ and $\Delta(\xi)$ follow the gap equations along the path.
Define $\mathcal E_\infty(\xi)$ as the limiting energy difference per site:
\begin{equation}
\mathcal E_\infty(\xi)=
\lim_{L\to\infty}\frac{E_0+E_{\bm k}-E_M}{L}
=-\mu(\xi)x(\xi)-\frac{\Delta(\xi)^2}{4g(\xi)}.
\end{equation}
Its derivative now includes the filling response:
\begin{equation}
\frac{\D\mathcal E_\infty}{\D\xi}
=-x\frac{\D\mu}{\D\xi}
-\mu\frac{\D x}{\D\xi}
-\frac1{4g}\frac{\D\Delta^2}{\D\xi}
+\frac{\Delta^2}{4g^2}\frac{\D g}{\D\xi}.
\label{eq:general-path-energy}
\end{equation}

Display the path dependence of the reduced action as
\begin{equation}
f(t;\xi)=\Phi(z_0(t;\xi),t;\xi)-x(\xi)\ln z_0(t;\xi).
\end{equation}
The function $\Phi$ is the same logarithmic integral as in the note.
Its explicit $\xi$ dependence includes the self-consistent BCS coefficients and energies.
The coupled limiting saddle equations are
\begin{equation}
\Phi_z(z_0,t_0;\xi)=\frac{x(\xi)}{z_0},
\qquad
\Phi_t(z_0,t_0;\xi)+i\mathcal E_\infty(\xi)=0.
\end{equation}
Equivalently, denote $f^{\prime}$ as the derivative with respect to $t$,
\begin{equation}
f'(t_0;\xi)+i\mathcal E_\infty(\xi)=0 ~~~~~~~~~~
K=\Phi_{zz}+\frac{x}{z_0^2}.
\end{equation}
Differentiating both saddle equations along the path gives
\begin{subequations}
\begin{equation}
K\frac{\D z_0}{\D\xi}
+\Phi_{zt}\frac{\D t_0}{\D\xi}
+\Phi_{z\xi}-\frac1{z_0}\frac{\D x}{\D\xi}=0,
\end{equation}
\begin{equation}
\Phi_{tz}\frac{\D z_0}{\D\xi}
+\Phi_{tt}\frac{\D t_0}{\D\xi}
+\Phi_{t\xi}+i\frac{\D\mathcal E_\infty}{\D\xi}=0.
\end{equation}
\end{subequations}
Subscripts denote partial derivatives.
In $\Phi_{z\xi}$ and $\Phi_{t\xi}$, the derivative with respect to $\xi$
holds $z$ and $t$ fixed.
Solving the coupled equations gives
\begin{equation}
\frac{\D t_0}{\D\xi} = -\frac{K(\Phi_{t\xi}+i\,\D\mathcal 
E_\infty/\D\xi) -\Phi_{tz}(\Phi_{z\xi}-z_0^{-1}\D x/\D\xi)}
{K\Phi_{tt}-\Phi_{zt}^{\,2}}
\label{eq:general-path-saddle-response}
\end{equation}
The term containing $\D x/\D\xi$ comes from number projection.
It cannot be omitted when the filling changes.

For $K\ne0$, elimination of the number saddle gives
\begin{equation}
f''=\Phi_{tt}-\frac{\Phi_{tz}\Phi_{zt}}{K},
\qquad
\partial_\xi f'
=\Phi_{t\xi}-\frac{\Phi_{tz}}{K}
\left(\Phi_{z\xi}-\frac1{z_0}\frac{\D x}{\D\xi}\right).
\end{equation}
Thus
\begin{equation}
\frac{\D t_0}{\D\xi}
=-\frac{\partial_\xi f'(t_0;\xi)+i\,\D\mathcal E_\infty/\D\xi}
{f''(t_0;\xi)}.
\end{equation}
If the coefficients are smooth and the coupled determinant is nonzero,
the saddle is locally smooth.
A boundary in the equilibrium phase diagram need not make this determinant vanish.

\paragraph{Case 1: singularity of derivative}

Suppose a contributing limiting saddle reaches $t_c$ at $\xi_c$ and satisfies
\begin{equation}
f'(t_c;\xi_c)+i\mathcal E_\infty(\xi_c)=0,\qquad
f''(t_c;\xi_c)=0,\qquad c=f'''(t_c;\xi_c)\ne0.
\end{equation}
Assume $K\ne0$ and that the following local expansion is valid.
Define $\delta\xi=\xi-\xi_c$ as the path displacement.
Use $\delta t=t_0-t_c$ as before.
Then
\begin{equation}
0=\frac c2(\delta t)^2+
\left[
\partial_\xi f'(t_c;\xi_c)
+i\left.\frac{\D\mathcal E_\infty}{\D\xi}\right|_{\xi_c}
\right]\delta\xi+\text{higher-order terms}.
\end{equation}
If the coefficient of $\delta\xi$ is nonzero,
\begin{equation}
t_0-t_c\simeq\pm\sqrt{-\frac{2}{c}\left[ \partial_\xi f'(t_c;\xi_c)
+i\left.\frac{\D\mathcal E_\infty}{\D\xi}\right|_{\xi_c}\right](\xi-\xi_c)}
\label{eq:general-path-cusp}
\end{equation}
Suppose this branch dominates on the regular side being considered.
If its prefactor contributes no finite additional phase derivative in the limit,
\begin{equation}
\tau_{\bm k}^{(\infty)}(\omega;\xi)=\operatorname{Re}t_0(\xi).
\end{equation}
Where the real square-root coefficient is nonzero,
\begin{equation}
\frac{\partial\tau_{\bm k}^{(\infty)}}{\partial\xi}
\ \propto\ |\xi-\xi_c|^{-1/2}.
\end{equation}
The delay has a finite one-sided limiting value and a singular slope.
This is a cusp, not a divergent delay value.
At the merging point, the joint saddle integral is still required.
If the numerator in Eq.~\eqref{eq:general-path-saddle-response} also vanishes,
the square-root conclusion need not hold.

\paragraph{Case 2: discontinuous jump of time delay}

Consider two distinct regular contributing branches $t_{0,1}$ and $t_{0,2}$.
Use their limiting actions per site,
\begin{equation}
\mathcal S_{\infty,j}(\xi)
=f(t_{0,j};\xi)+i\mathcal E_\infty(\xi)t_{0,j},
\qquad j=1,2.
\end{equation}
At the crossing, require equal real actions:
\begin{equation}
\operatorname{Re}[\mathcal S_{\infty,1}(\xi_c)
-\mathcal S_{\infty,2}(\xi_c)]=0.
\end{equation}
A simple change of their ordering requires
\begin{equation}
\left.\frac{\D}{\D\xi}
\operatorname{Re}[\mathcal S_{\infty,1}-\mathcal S_{\infty,2}]
\right|_{\xi_c}\ne0.
\end{equation}
The envelope relation now contains the filling derivative:
\begin{equation}
\frac{\D\mathcal S_{\infty,j}}{\D\xi}
=\Phi_\xi(z_0(t_{0,j}),t_{0,j};\xi)
-\frac{\D x}{\D\xi}\ln z_0(t_{0,j};\xi)
+i\frac{\D\mathcal E_\infty}{\D\xi}t_{0,j}.
\end{equation}
Suppose branch 1 dominates below $\xi_c$ and branch 2 above it.
Both must be accessible from the original contours.
Other saddles or endpoint contributions must not dominate them.
For regular prefactors, the limiting delay has a jump if
\begin{equation}
\lim_{\xi\to\xi_c^+}\tau_{\bm k}^{(\infty)}
-\lim_{\xi\to\xi_c^-}\tau_{\bm k}^{(\infty)}
=\operatorname{Re}[t_{0,2}(\xi_c)-t_{0,1}(\xi_c)]\ne0
\label{eq:general-path-jump}
\end{equation}
Both limits are taken at the same fixed $\omega$ and $\eta$.
An exchange of dominant actions with equal real saddle times does not give a delay jump.
At the crossing itself, the coherent sum of the two contributions must be retained.

\paragraph{Examples at two boundary lines}

On the Moore--Read line,
\begin{equation}
\mu=\frac{\Delta^2}{4}>0,\qquad
x(\xi_c)=1-\frac1{g(\xi_c)},\qquad
\mathcal E_\infty(\xi_c)=-\mu(\xi_c).
\end{equation}
A transverse crossing satisfies
\begin{equation}
\left[\frac{\D x}{\D\xi}
-\frac1{g^2}\frac{\D g}{\D\xi}\right]_{\xi_c}\ne0.
\end{equation}
The time-saddle equation at the crossing is
\begin{equation}
f'(t_0;\xi_c)=i\mu(\xi_c).
\end{equation}
For a cusp, this must be supplemented by $f''(t_c;\xi_c)=0$ and the
nonzero coefficient in Eq.~\eqref{eq:general-path-cusp}.
For a jump, the two candidate actions are
\begin{equation}
\mathcal S_{\infty,j}(\xi_c)
=f(t_{0,j};\xi_c)-i\mu(\xi_c)t_{0,j}.
\end{equation}
Their dominance must exchange, and their real saddle times must differ.

On the Read--Green line,
\begin{equation}
\mu=0,\qquad
2x(\xi_c)=1-\frac1{g(\xi_c)},\qquad
\mathcal E_\infty(\xi_c)=-\frac{\Delta(\xi_c)^2}{4g(\xi_c)}.
\end{equation}
A transverse crossing satisfies
\begin{equation}
\left[2\frac{\D x}{\D\xi}
-\frac1{g^2}\frac{\D g}{\D\xi}\right]_{\xi_c}\ne0.
\end{equation}
The corresponding saddle equation and branch actions are
\begin{equation}
f'(t_0;\xi_c)=\frac{i\Delta(\xi_c)^2}{4g(\xi_c)},
\qquad
\mathcal S_{\infty,j}(\xi_c)
=f(t_{0,j};\xi_c)-\frac{i\Delta(\xi_c)^2}{4g(\xi_c)}t_{0,j}.
\end{equation}
The same cusp and jump tests apply.
The gapless endpoint on this line requires a separate check of the derivatives
used in the Taylor expansion.

\subsubsection{Physical interpretations}

Below are the physical interpretations of two features of Wigner time delay $\tau_{\bm{k}}(\omega)$ near phase boundary lines.
\begin{itemize}
\item Singularity of derivative $\partial_\xi \tau_{\bm{k}}(\omega, \xi)$: 
	The response of wavefunction's phase with respect to the excitation through transition amplitude. 
\item Discontinuous jump of $\tau_{\bm{k}}(\omega, \xi)$:
	The interference from the exchange of dominant contributions in transition amplitudes.
\end{itemize}

\paragraph{Case 1: singularity of derivative}

In the transition amplitude $D_{\bm{k}}(\omega)$, 
each excited spectator pair contributes a phase $e^{2iE_{\bm p}t}$.
Their phase are combined in 
\begin{equation}
\prod_{\bm p\ne\bm k}
[A_{\bm p}(z)+B_{\bm p}(z)e^{2iE_{\bm p}t}].
\end{equation}
The functions $\Phi$ and $f(t;\xi)$ collect this coherent configuration sum.
The saddle point characterizes a collective phase response,
\begin{equation}
f'(t_0;\xi)+i\mathcal E_\infty(\xi)=0.
\end{equation}
The response equation is
\begin{equation}
\frac{\D t_0}{\D\xi} =-\frac{\partial_\xi f'(t_0;\xi)+i\,
\D\mathcal E_\infty/\D\xi}{f''(t_0;\xi)}.
\end{equation}
When the reduced curvature becomes small, the stationary time is weakly constrained.
A small parameter change can then move it strongly.
If the curvature vanishes under the conditions of Eq.~\eqref{eq:general-path-cusp},
the saddle has a square-root branch.
The singular slope expresses a large sensitivity of the coherent phase response.
Projection supplies configuration mixing that can support this mechanism.
It does not guarantee that the required curvature vanishes.

\paragraph{Case 2: discontinuous jump}

Consider a simple case, where two competing saddle contributions,
\begin{equation}
D_{\bm k}\simeq D_{\bm k,1}+D_{\bm k,2}.
\end{equation}
On regular branches, neglecting subleading prefactor derivatives,
\begin{equation}
\partial_\omega D_{\bm k,j}\simeq i t_{0,j}D_{\bm k,j},
\qquad j=1,2.
\end{equation}
Each saddle represents a collective contribution of many configurations,
\begin{equation}
\tau_{\bm k}\simeq\operatorname{Re}
\frac{t_{0,1}D_{\bm k,1}+t_{0,2}D_{\bm k,2}}
{D_{\bm k,1}+D_{\bm k,2}}.
\end{equation}
If dominance exchanges in the thermodynamic limit, the selected delay can change
from $\operatorname{Re}t_{0,1}$ to $\operatorname{Re}t_{0,2}$.
Unequal one-sided real times give the jump in Eq.~\eqref{eq:general-path-jump}.

\medskip
To see how the exchange produces a jump, keep $\omega$ and $\eta>0$ fixed.
As an illustration, suppose the two contributions have the same phase
at the chosen frequency and their ratio is
\begin{equation}
\frac{D_{\bm k,2}}{D_{\bm k,1}}
=e^{L a(\xi-\xi_c)},\qquad a>0.
\label{eq:jump-crossover-ratio}
\end{equation}
Here $a$ gives the rate at which the real action per site of contribution~2
increases relative to that of contribution~1 along the path $\xi$.
The assumption on the phase applies at the chosen frequency.
The two frequency derivatives of the phase can still differ.

Treating $t_{0,1}$ and $t_{0,2}$ as approximately constant near $\xi_c$,
the two-contribution expression for the delay becomes
\begin{equation}
\tau_{\bm k}\simeq
\frac{\operatorname{Re}t_{0,1}
+e^{L a(\xi-\xi_c)}\operatorname{Re}t_{0,2}}
{1+e^{L a(\xi-\xi_c)}}.
\label{eq:jump-crossover-delay}
\end{equation}
For finite $L$, this is a smooth crossover.
Both contributions have comparable magnitudes over a range of order
$1/(La)$ in $\xi$. This range shrinks as $L$ increases:
\begin{equation}
\lim_{L\to\infty} e^{La(\xi-\xi_c)}
=
\begin{cases}
0,&\xi<\xi_c,\\
+\infty,&\xi>\xi_c.
\end{cases}
\end{equation}
At fixed $\xi<\xi_c$ the limiting delay is set by
$\operatorname{Re}t_{0,1}$, and at fixed $\xi>\xi_c$ it is set by
$\operatorname{Re}t_{0,2}$. Taking the one-sided limits after
$L\to\infty$ gives
\begin{equation}
\lim_{\xi\to\xi_c^+}\lim_{L\to\infty}\tau_{\bm k}
-
\lim_{\xi\to\xi_c^-}\lim_{L\to\infty}\tau_{\bm k}
=\operatorname{Re}\bigl[t_{0,2}(\xi_c)-t_{0,1}(\xi_c)\bigr].
\end{equation}
Each saddle time may be smooth across the boundary. The jump occurs
when the selected contribution changes and the two limiting real
times differ.

For the full complex ratio, the relative phase affects the crossover
and can affect the delay at $\xi_c$. The coherent sum is needed wherever
the contributions are comparable and $D_{\bm k}\ne0$. To establish a jump
on either BCS boundary, one must show that two contributing saddles exchange
dominance there and control both the amplitude and its frequency derivative
on their respective sides. Equation~\eqref{eq:jump-crossover-delay} illustrates
this mechanism.

\bibliography{main}